\documentclass[preprintnumbers,preprint,showpacs,prd]{revtex4}
\usepackage{graphicx}
\input psfig.sty
\input epsf.sty
\newcommand{\be}{\begin{equation}}
\newcommand{\ee}{\end{equation}}
\newcommand{\ba}{\begin{eqnarray}}
\newcommand{\ea}{\end{eqnarray}}

\begin{document}

\preprint{BU-HEPP-04-02}
%\date{\today}
\title{Electric Polarizability of Neutral Hadrons from Lattice QCD}
\author{Joe Christensen}
\affiliation{Physics Department, McMurry University, Abilene, TX,
79697}
\author{Walter Wilcox}
\affiliation{Department of Physics, Baylor University, Waco, TX 76798-7316}
\author{Frank X.\ Lee$^{a,b}$ and Leming Zhou$^{a}$}
\affiliation{$^{a}$Center for Nuclear Studies, George Washington
University, Washington DC, 20052 \\
$^{b}$Jefferson Lab, 12000 Jefferson Avenue, Newport News, VA, 23606}

\begin{abstract}
By simulating a uniform electric field on a lattice and measuring
the change in the rest mass, we calculate the electric
polarizability of neutral mesons and baryons using the methods of
quenched lattice QCD. Specifically, we measure the electric
polarizability coefficient from the quadratic response to the
electric field for 10 particles: the vector mesons $\rho^0$ and
$K^{*0}$; the octet baryons n, $\Sigma^0$, $\Lambda_{o}^{0}$,
$\Lambda_{s}^{0}$, and $\Xi^0$; and the decouplet baryons
$\Delta^0$, $\Sigma^{*0}$, and $\Xi^{*0}$. Independent
calculations using two fermion actions were done for consistency
and comparison purposes. One calculation uses Wilson fermions with
a lattice spacing of $a=0.10\,$fm. The other uses tadpole improved
L\"usher-Weiss gauge fields and clover quark action with a lattice
spacing $a=0.17\,$fm. Our results for neutron electric
polarizability are compared to experiment.
\end{abstract}


\maketitle

\section{Introduction and Review}

Electric and magnetic polarizabilities characterize the rigidity
of both charged and uncharged hadrons in external fields and are
important fundamental properties of particles. In particular, the
electric polarizability of a hadron characterizes the reaction of
quarks to a weak external electric field and can be measured by
experiment via Compton scattering. In this paper we describe a
lattice quantum chromodynamics (QCD) calculation of neutral hadron
electric polarizabilities using an external field method. The goal
of Monte Carlo lattice QCD is to extract fundamental, measurable
quantities directly from the theory without model assumptions.
Learning about such aspects of particles tests our understanding
and formulation of the underlying theory and makes new aspects and
predictions of the theory subject to experimental verification.
Such calculations can give insights on the internal structure of
hadrons and the applicability of chiral perturbation theory to
various low energy aspects of lattice QCD.

Conceptually, electric and magnetic fields, even from a single
photon, will distort the shape of a hadron, thereby affecting the
internal energy and thus the mass. The electric and magnetic
polarizabilities are defined as the coefficients of the quadratic
electric and magnetic field terms in the mass shift formula
($\hbar=c=1$ gaussian units):
\begin{equation}
\Delta m = -\frac{1}{2} \, \alpha \vec{E}^2 - \frac{1}{2} \,\beta
\vec{B}^2,\label{eq:empol}
\end{equation}
where ${\alpha}$ $(\beta)$ is the electric (magnetic)
polarizability that compares to experiment. When the terms in
Eq.(\ref{eq:empol}) are viewed as the non-relativistic interaction
Hamiltonian, one obtains the polarized on-shell Compton scattering
cross section for neutral particles:

\begin{equation}
\frac{d\sigma}{d\Omega}
=\frac{1}{(4\pi)^2}\left(\frac{\omega_2}{\omega_1}\right)|\,\alpha\,
\omega_1\,\omega_2\,
(\hat{\epsilon_1}\cdot\hat{\epsilon_2}^*) + \beta\,
(\hat{\epsilon_1}\times\vec{k}_1)\cdot(\hat{\epsilon_2}^*\times\vec{k}_2)
\,|^2.\label{eq:cross}
\end{equation}
$\omega_{1,2}$ and $\vec{k}_{1,2}$ are the initial, final photon
angular frequencies and wave vectors and $\hat{\epsilon}_{1,2}$
are the polarization vectors. (The quantity $\omega_2/\omega_1$ is
a recoil factor and can be ignored nonrelativistically.) This
allows hadron polarizabilities to be measured in scattering
experiments. (For charged spinless particles one must also add the
Thompson scattering amplitude from the particle's charge and, if
virtual, a charge radius term before calculating the cross
section. See Eq.(11) of \cite{wil1}.) Compilations of experimental
results for neutron electric polarizability have been given in
\cite{chr1,Zhou:2004}. Although the polarizabilities of the other
particles investigated here have not been measured, we hope that
the comparison of the results from the various types of mesons and
baryons investigated will give insights on their relative rigidity
and structure.

The lattice calculation of electric polarizabilities began with
the paper by Fiebig {\it et al.\/}~\cite{Fiebig:1989en} using the
staggered fermion formulation of lattice quarks. An external
electric field was simulated on the lattice and mass shifts
measured directly for the neutron and neutral pion. Although the
simulation errors were large, the neutron electric polarizability
extracted there is now seen to be in remarkable agreement with
recent experiments. Lattice four point function techniques have
also been designed to extract neutral or charged particle electric
polarizabilities~\cite{wil1} (chiral symmetry can sometimes be
used to reduce the calculation to two point
functions~\cite{wil2}), but these methods are more difficult to
carry out on the lattice. Early results of the present study have
been reported in \cite{chr1}. See~\cite{Zhou:2002} for preliminary
results of a companion calculation of the magnetic polarizability
of both charged and uncharged hadrons.

\section{Lattice Details}

The clover part~\cite{lush} of this calculation uses the
tadpole-improved clover action with coupling constant 
$c_{SW}=1/u_0^4$, where $u_0^4$
is the average plaquette. We use the zero-loop, tadpole
improved L\"usher-Weiss gauge field action on a quenched
$12^3\times 24$ lattice with $\beta=7.26$ ($a_{\rm
clover}=0.17\,{\rm fm}$). In both Wilson and clover cases the
gauge field was thermalized by 10,000 sweeps (quasi-heatbath with
overrelaxation) and then saved every 1000. We have used 100
configurations in the clover case. In the Wilson case the lattice
is $24^4$ and we used 109 configurations with standard gauge field
action with $\beta=6.0$ ($a_{\rm Wilson}=0.1\,{\rm
fm}$)~\cite{gock}. Our quark propagator time origin, $t=0$, was
chosen to be the third lattice time step for clover fermions and
the second time step in the Wilson case. Point quark sources were
constructed for the zero momentum quark propagators. The standard
particle interpolation fields for the octet~\cite{terry1}
(non-relativistically non-vanishing) and decouplet~\cite{terry2}
baryons were used. Periodic boundary conditions in the spatial
directions and free or Dirichlet boundary conditions for the time
links on the lattice time edges were used.

As is usual in lattice calculations, the mass of a hadron can be
calculated from the exponential time decay of a correlation
function. By calculating the ratio of the correlation function in
the field to that without the field, we have a ratio of
exponentials which decays at the rate of the mass difference.
According to Eq.(\ref{eq:empol}), the electric mass shifts should
be parabolic, negative mass shifts giving positive polarizability
coefficients and positive shifts giving negative coefficients. We
use four values of the electric field to establish the parabola.
Following the technique in \cite{Fiebig:1989en}, we will average
the correlation functions over $\vec{E}$ and $-\vec{E}$ in order
to remove the spurious linear term in the parabolic fits. We
include the uniform static E-field as a phase on the gauge links
in a particular direction (with fermion charge $q=Qe$):
\begin{equation}
e^{iaqA} = e^{i(a^2qE)(x_4/a)} = e^{i \eta\tau} \rightarrow
(1+i\eta \tau).\label{eq:efield}
\end{equation}
There is a discontinuity in the electric field at the decoupled
lattice time boundary under this formulation, which should not be
a problem as all of our correlation functions are measured far
from the time edges of our lattices. Since the electric field is
linearized in the continuum, we used the linearized form on the
lattice. Fiebig {\it et al.\/} found no significant difference
between the exponential and linearized formats for similar
electric fields.

On the lattice we have an exact SU(2) isospin symmetry. This
symmetry is broken by the different electric charges on the u- and
d-quarks in the presence of the external electric field.
Consequently the $\pi^0$ and $\rho^0$ $I=$1 particles can mix with
$I=0$ glueballs and disconnected quark loops (self-contractions of
the interpolation fields) can propagate these particles.
Disconnected loop methods have been considerably improved in
recent years~\cite{defl}; however, calculation of these diagrams
is still extremely time consuming. In this paper, we will ignore
the effect of the disconnected loops.

Including the static electric field as a phase on the links
affects the Wilson term, but not the clover loops. This is clear
since the conserved vector current derived by the Noether
procedure is identical for the clover and Wilson actions. In units
of $10^{-3}e^{-1}a^{-2}$, the electric field took the values
$\pm 1.08$, $\pm 2.16$, $\pm 4.32$, and $\pm 8.64$ via the parameter
$\eta=a^2QeE$ in Eq.~(\ref{eq:efield}) for both the clover and
Wilson cases. (The $\eta$ values were $\pm 0.00036, \pm 0.00072,
\pm 0.00144, \pm 0.00288$, and $\pm 0.00576$, the same as in
Ref.~\cite{Fiebig:1989en}.) In conventional units the smallest
electric field is approximately $7.4\times 10^{21}$ N/C in the
clover case [$(.17/.1)^2=2.35$ times larger in the Wilson case],
which is about the same electric field strength $.26\,{\rm fm}$
from a d-quark. Although these are huge fields, we will
nevertheless see that the lattice mass shifts are small and that
there is no evidence of $E^4$ or higher terms in the electric
field fits. We will in fact check that a drastic reduction in the
field does not change the polarizability coefficients.

We did the calculation with both Wilson and clover fermions in
order to test the consistency of our results as well as to set
benchmarks for future polarizability calculations. We do not
expect that the two formulations will agree with one another
throughout the mass range investigated. As in any lattice
calculation, there are many sources of systematic error, including
quenching, finite volume, and finite size errors. The finite volume errors for
the Wilson case ($L\simeq 2.4$ fm) should be slightly smaller than
for clover ($L\simeq 2$ fm); however, discretization errors should
be smaller for the clover case. In addition, to achieve physical
results, a chiral extrapolation to physical quark masses is
necessary. The quenched chiral regime has been estimated to extend
to pion masses of only about $300$ MeV~\cite{kehfei} for the
pseudoscalar decay constant, $f_P$, and the ratio $m_{\pi}^2/m$,
where $m$ is the quark mass, although this does not exclude the
range being larger for other quantities. Our pion masses are
almost certainly too large to get into the chiral regime and we do
not attempt a chiral extrapolation here. However, we would expect
the agreement between the two calculations to improve at our lower
pion masses, which is what is seen. Our pion masses here range
from about 1 GeV to about 500 MeV; we achieve the smallest pion
mass in the case of clover fermions, 483 MeV.

With $\kappa_{\rm cr} = 0.1232(1)$ in the clover case, we used six
values of $\kappa = 0.1182$, $0.1194$, $0.1201$, $0.1209$,
$0.1214$, and $0.1219$, which roughly correspond to $m_q\sim 200$,
$150$, $120$, $90$, $70$, and $50$ MeV. The critical value of
$\kappa$ for the Wilson case is
$0.157096(28)^{+33}_{-9}$~\cite{iwasaki}. The values of $\kappa$
used in the Wilson calculation were
$0.1515,0.1525,0.1535,0.1540,0.15454$, and $0.1555$, which correspond to
approximate quark masses of $m_q\sim 232$, $189$, $147$, $126$,
$106$, and $65$ MeV. We choose $\kappa=0.1201$ to represent clover
strange quarks and $\kappa=0.1540$ to represent Wilson strange
quarks. We used multi-mass BiCGStab as the inversion algorithm for
both the clover and Wilson cases. The number of quark inversions
per gauge field and quark mass was 11: 10 values of $\eta$ in
Eq.(\ref{eq:efield}) associated with 4 nonzero electric fields for
both the u and d-quarks, plus the zero field inversion. Our Wilson
and clover calculations were done completely independently from
one another.

\section{Results}

\begin{figure}
\begin{center}
\includegraphics[width=4.50in ]{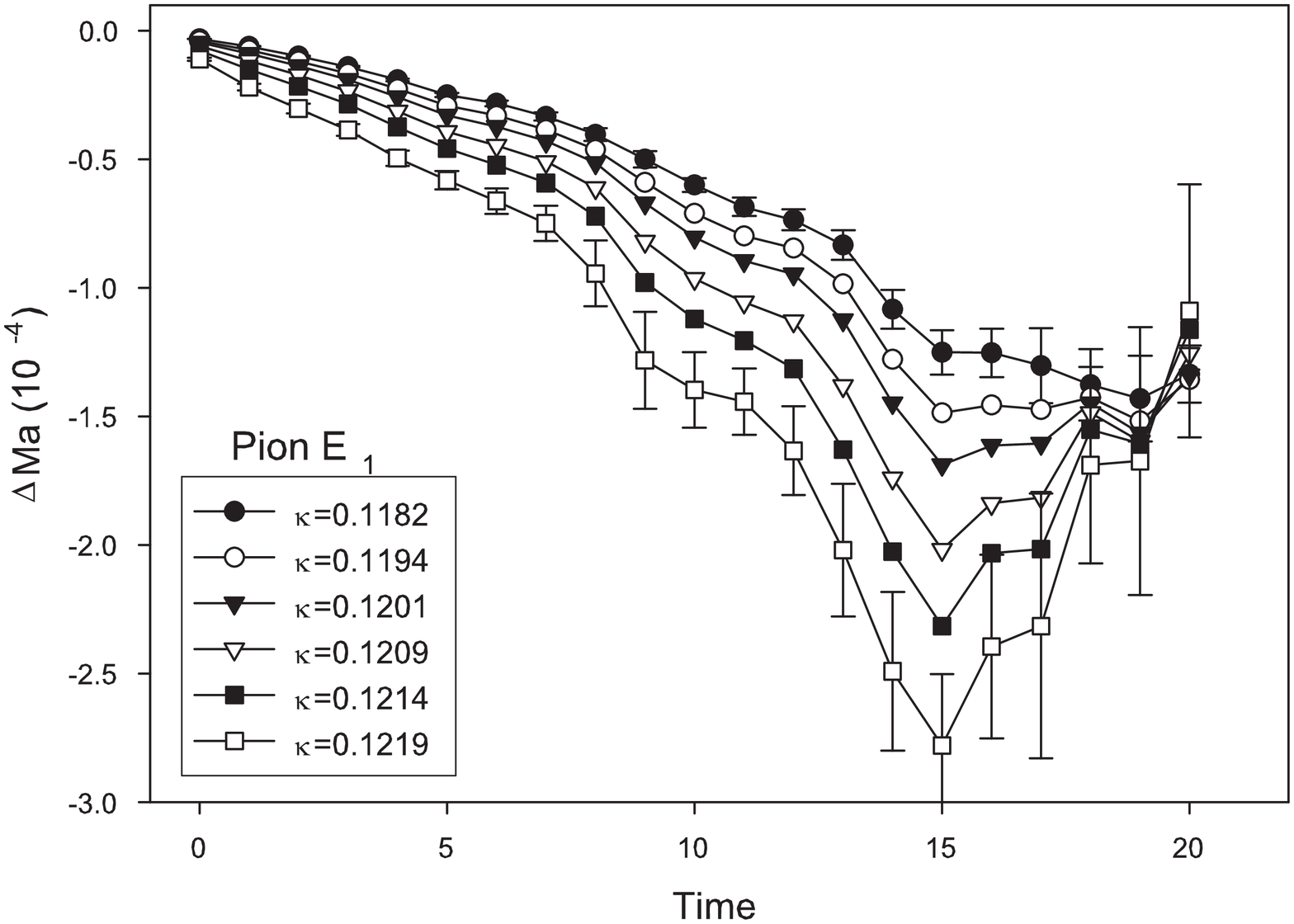}
\caption{The effective lattice mass shift, $\Delta Ma$, for the
$\pi^0$ for the six quark masses in the clover study for the
smallest electric field value. Error bars are shown only on the
$\kappa=0.1182,0.1219$ values.} \label{pion:mass-shift}
\vspace{1cm}
\includegraphics[width=3.1in, angle=90]{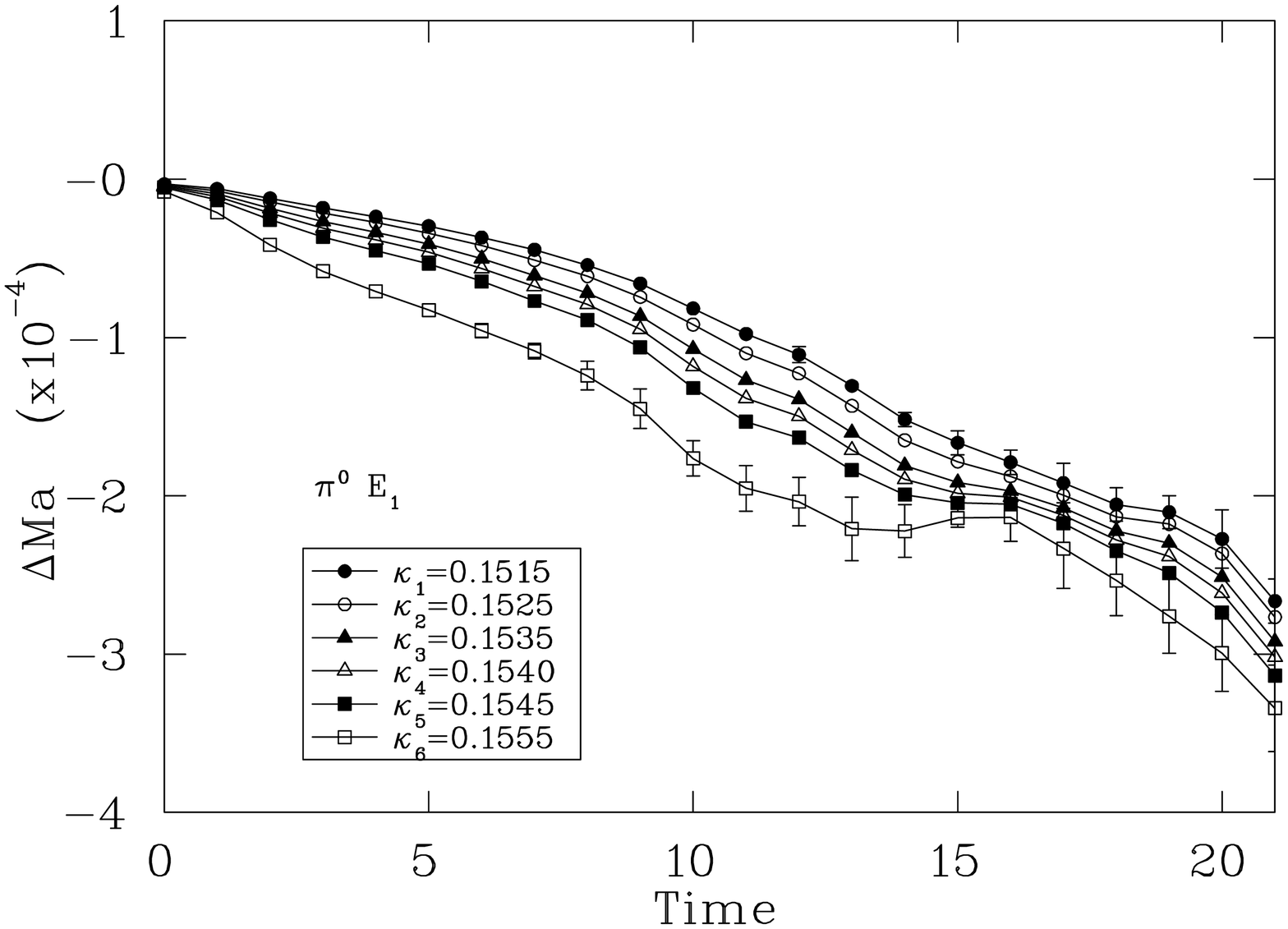}
\caption{The effective lattice mass shift, $\Delta Ma$, for the
$\pi^0$ for the six quark masses in the Wilson study for the
smallest electric field value. Error bars are shown only on the
$\kappa=0.1515, 0.1555$ values.} \label{pion:Wmass-shift}
\end{center}
\end{figure}
\begin{figure}
\begin{center}
\includegraphics[width=5.00in ]{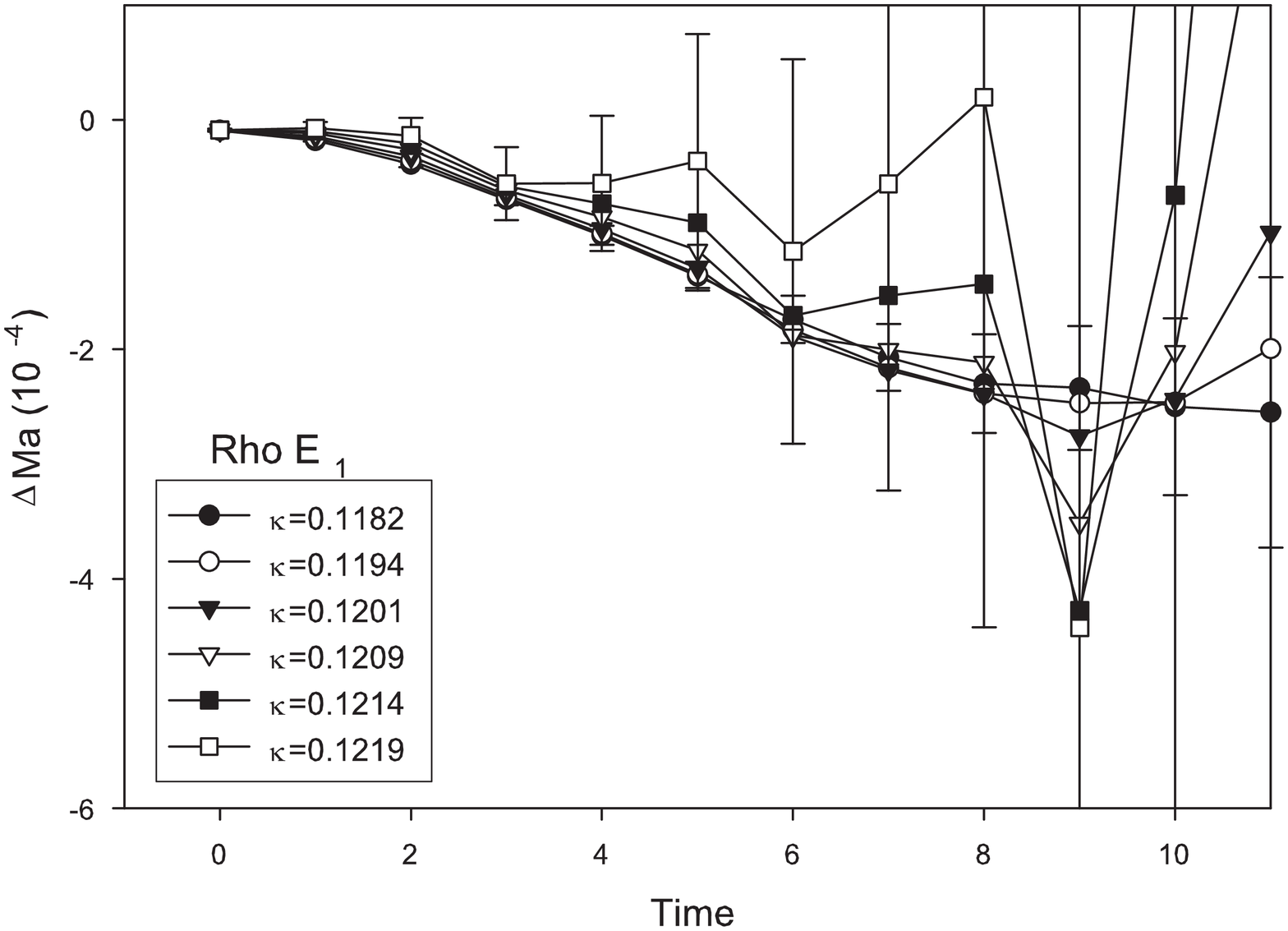}
\caption{Same as Fig.~1 (clover case) except for the $\rho^0$.}
\label{rho:mass-shift}
\vspace{1cm}
\includegraphics[width=3.5in, angle=90 ]{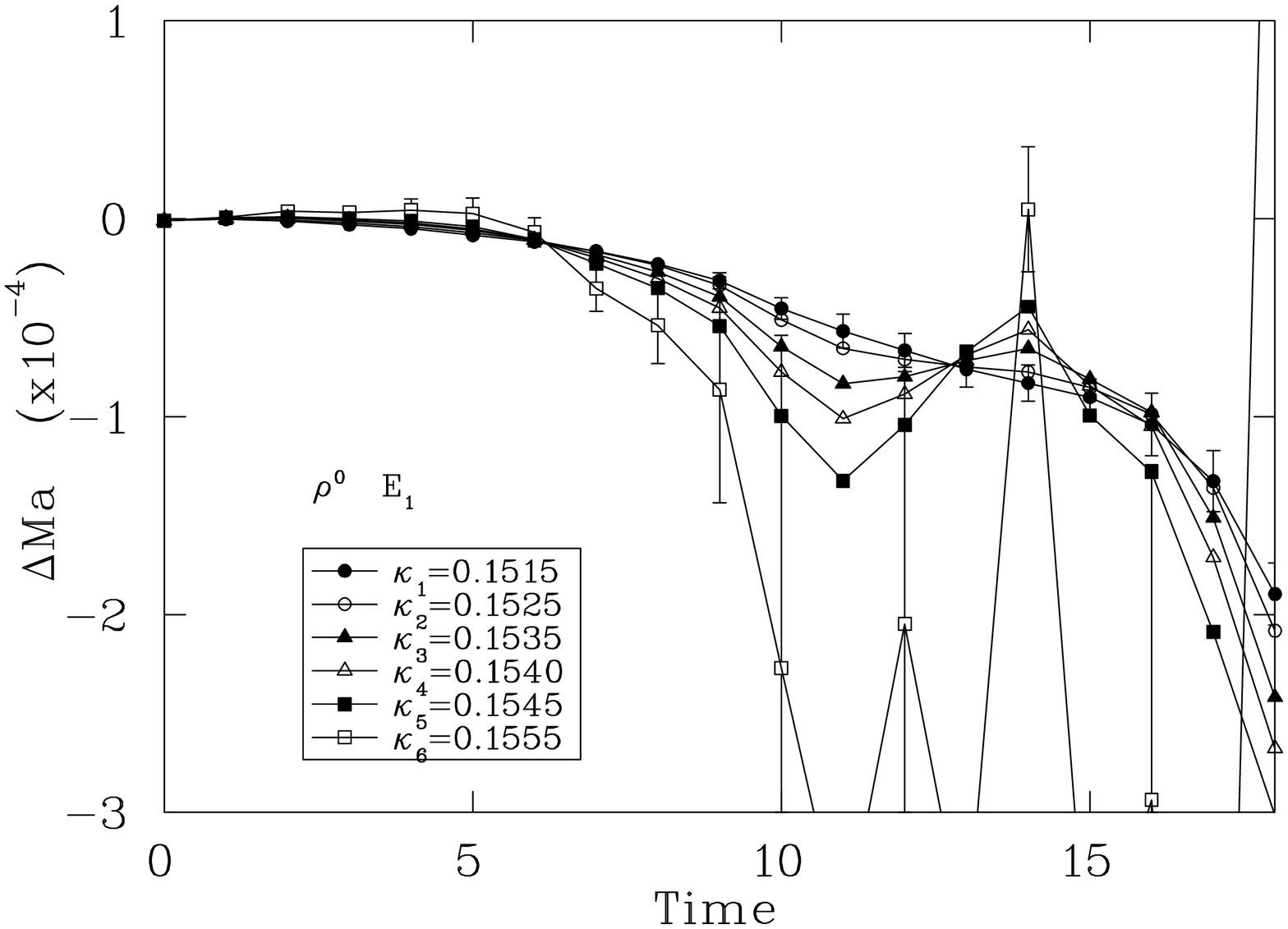}
\caption{Same as Fig.~2 (Wilson case) except for the $\rho^0$.}
\label{rho:Wmass-shift}
\end{center}
\end{figure}
\begin{figure}
\begin{center}
\includegraphics[width=5.00in ]{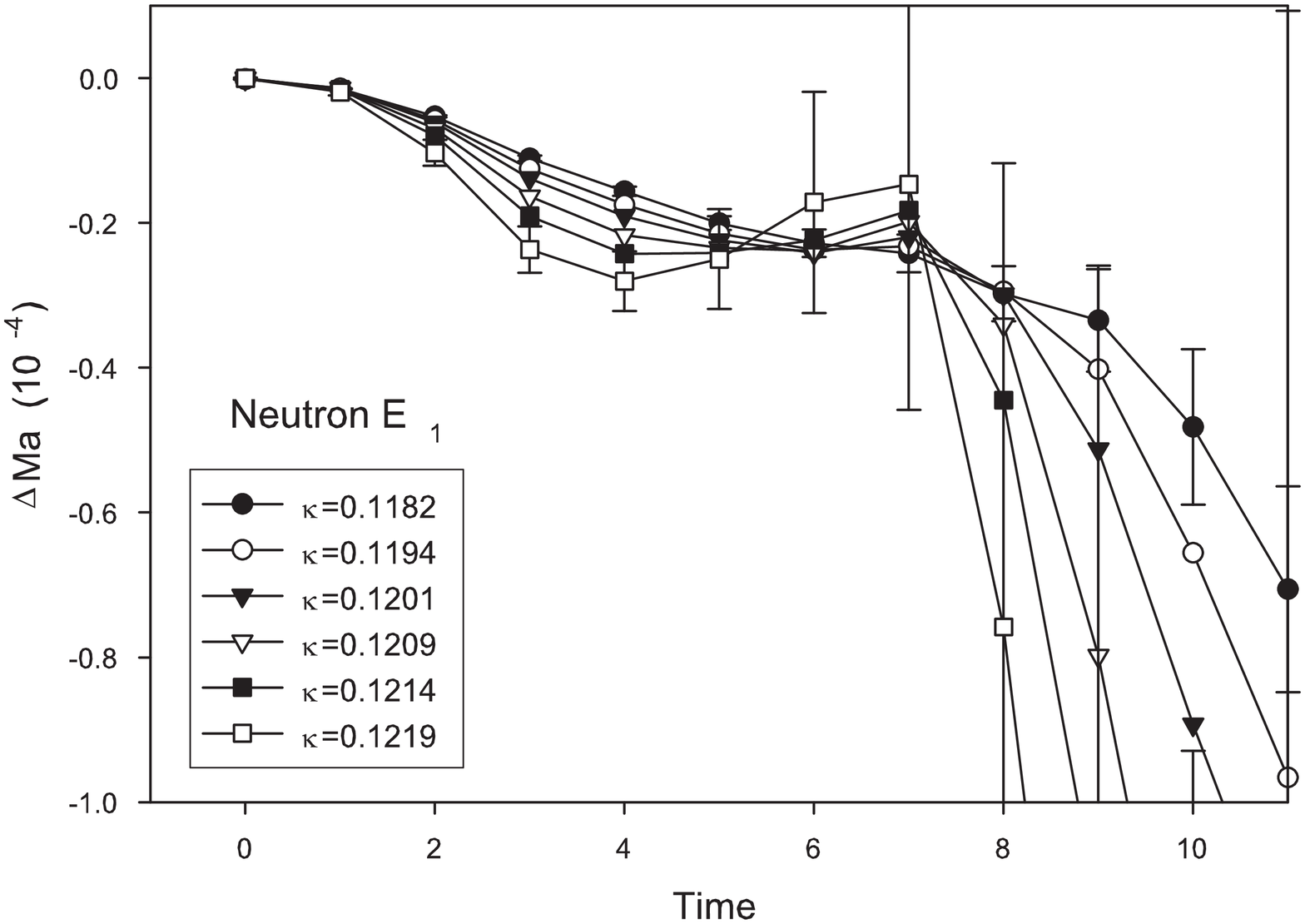}
\caption{Same as Fig.~1 (clover case) except for the neutron.}
\label{neutron:mass-shift}
\vspace{1cm}
\includegraphics[width=3.5in, angle=90 ]{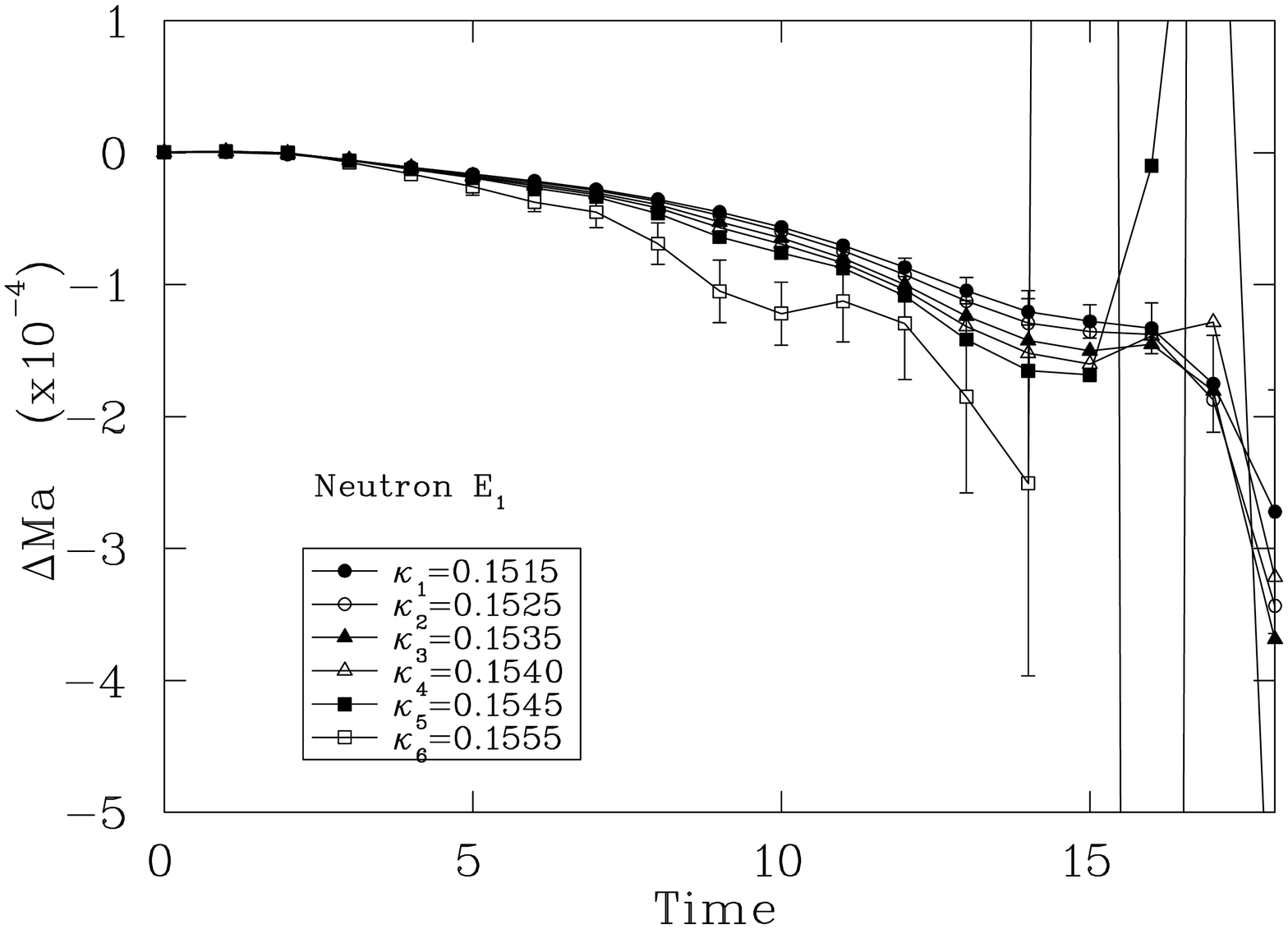}
\caption{Same as Fig.~2 (Wilson case) except for the neutron.}
\label{neutron:Wmass-shift}
\end{center}
\end{figure}
\begin{figure}
\begin{center}
\includegraphics[width=5.00in ]{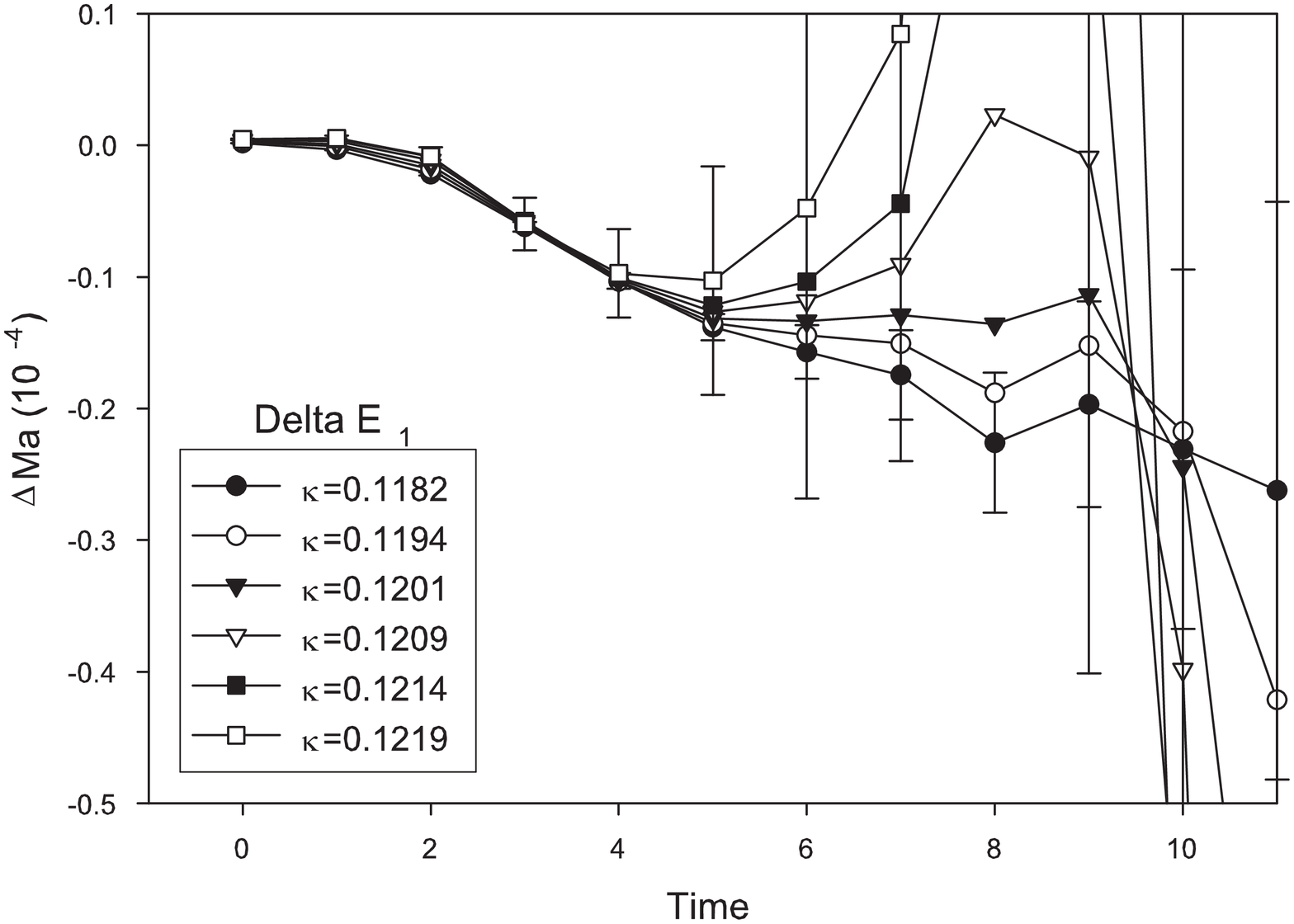}
\caption{Same as Fig.~1 (clover case) except for the $\Delta^0$.}
\label{delta:mass-shift} \vspace{1cm}
\includegraphics[width=3.5in, angle=90 ]{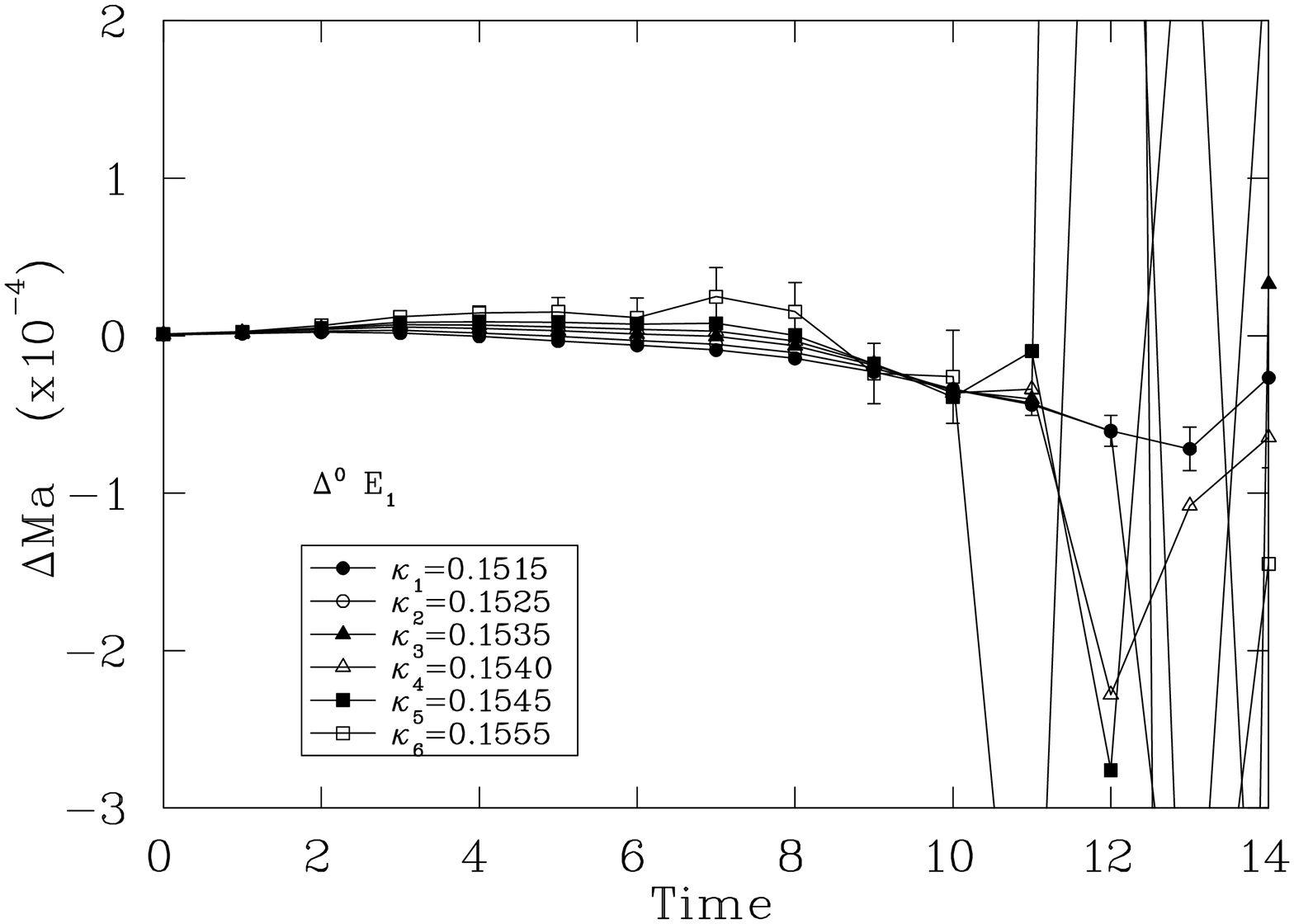}
\caption{Same as Fig.~2 (Wilson case) except for the $\Delta^0$.}
\label{delta:Wmass-shift}
\end{center}
\end{figure}

We start with an examination of the effective mass shifts of the
four non-strange particles in this study, $\pi^0$, $\rho^0$,
neutron, and $\Delta^0$, as a function of lattice time; see
Figs.~\ref{pion:mass-shift}-\ref{delta:Wmass-shift}. The effective
mass shift for each particle is defined to be
\be
R(t) = \frac{G_{E}(t)}{G_0(t)}, \ee
\be
\Delta Ma\,(t)\equiv \ln\, (\frac{R(t)}{R(t+1)}).\label{ratio} \ee
where $G_0(t)$ is the particle propagator without an electric
field and $G_{E}(t)$ is the value with the field. The notation
$\Delta Ma\,(t)$ indicates this value is being associated with the
time point $t$ (measured from the hadron source) in the graphs.
These figures are used to guide us in the choice of optimal
propagator time points in the fits. Just as single exponential
behavior should emerge for each particle channel in Euclidean
time, the ratio of particle propagators must also become single
exponential. This means, from Eq.(\ref{ratio}), that the effective
mass shift plot should become time independent. As in most lattice
simulations, particle propagators become increasing noisy as
lattice time increases, so the mass shift data will eventually
become dominated by statistical errors at large time separations.
We will examine the results at the lowest electric field value;
the effective mass shifts at larger fields we will soon see is
simply scaled with the $E^2$ value.

Figs.~\ref{pion:mass-shift}, \ref{rho:mass-shift},
\ref{neutron:mass-shift}, and \ref{delta:mass-shift} give the
non-strange hadron mass shifts for six quark masses of clover fermions;
Figs.~\ref{pion:Wmass-shift}, \ref{rho:Wmass-shift},
\ref{neutron:Wmass-shift}, and \ref{delta:Wmass-shift} give
similar shifts for six Wilson masses. Error bars computed with the
jackknife technique are shown on the largest and smallest quark
mass results to give an idea of the statistical errors. We show
only those lattice time points which we feel have meaningful Monte
Carlo values in each of these graphs for clarity. Our pion result,
Figs.~\ref{pion:mass-shift} and \ref{pion:Wmass-shift}, is
surprising. We never do see the expected time independence of the
mass shift for any value of the quark mass. Our results in
Figs.~\ref{pion:mass-shift} and \ref{pion:Wmass-shift} are plotted
out to the final time step (20 in the clover case, 21 in the
Wilson) to show the complete time history. There is apparently a
dip in the time behavior for clover fermions near $t=15$, but our
error bars are too large at this point to confirm this as a
plateau. The Wilson case also has no convincing plateau. We note
that Fiebig {\it et al.\/}~\cite{Fiebig:1989en} isolated a small
signal for the neutral pion in their calculations. They used
staggered fermions, which have a reminant exact chiral symmetry.
Of course the chiral symmetry is broken for both Wilson and clover
fermions, and this could be the crucial difference in the
calculations. It is possible that the neglect of the disconnected diagrams
could be responsible for the bad pion behavior. The prediction from
chiral perturbation theory is that $\alpha_{\pi^0}$ should
be small and negative: $\alpha_{\pi^0}= -0.35\pm .1
\times 10^{-4}$~fm$^3$~\cite{bell}.

\begin{figure}
\begin{center}
\includegraphics[width=6.00in]{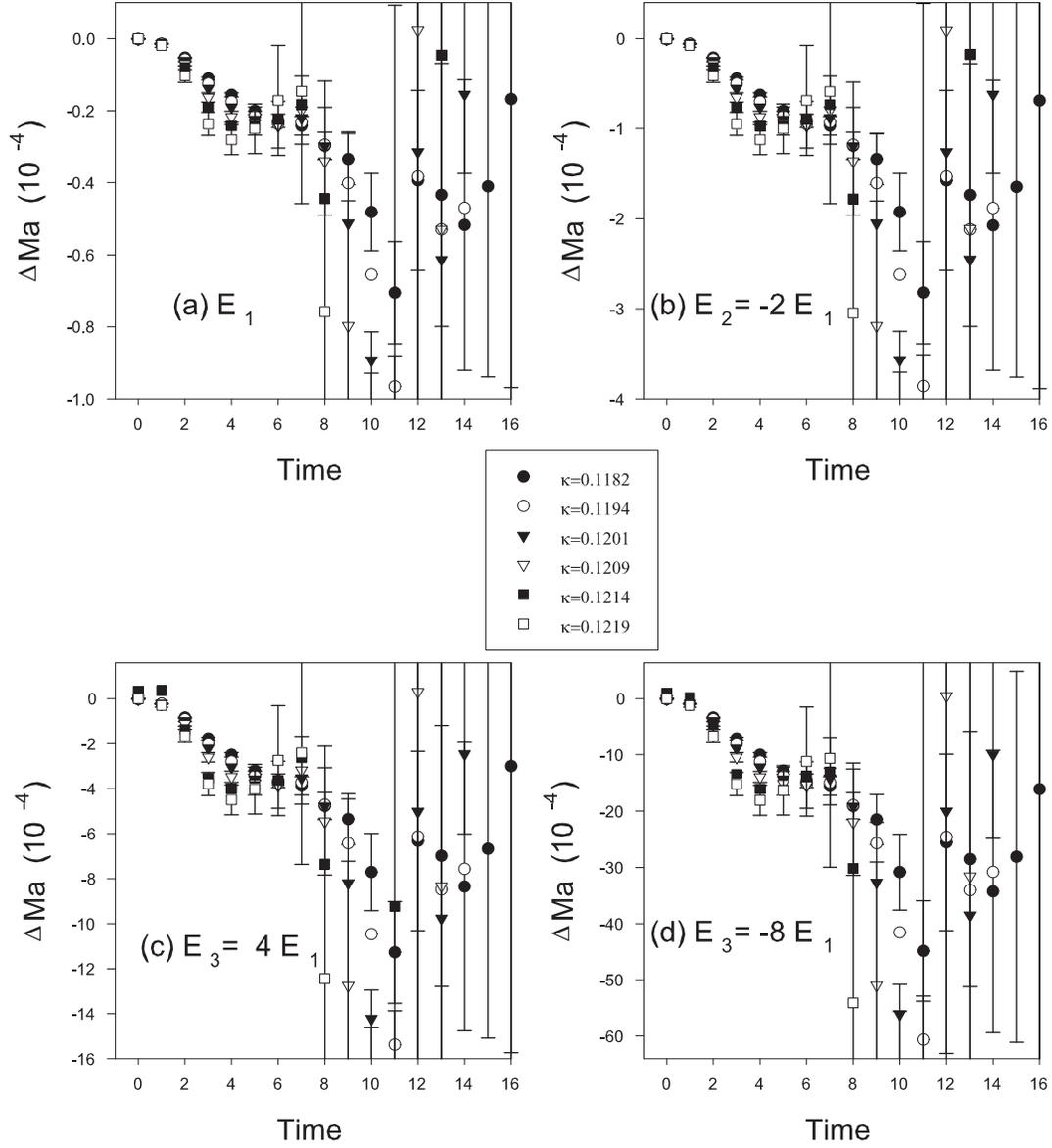}
\caption{The neutron effective lattice mass shifts, $\Delta Ma$,
evaluated for 4 values of the external electric field,
E$_{1}=-1.08 \times 10^{-3}e^{-1}a^{-2}$, -2E$_{1}$, 4E$_{1}$, and
-8E$_{1}$ in (a)-(d) respectively, for six quark mass values.
Error bars shown only on the $\kappa=0.1182,0.1219$ values. The
vertical axis scale increases by a factor of 4 in each case,
resulting in strikingly similar figures.} \label{neutron:amaze}
\end{center}
\end{figure}

The other particle channels studied behave conventionally.
Figs.~\ref{rho:mass-shift} and \ref{rho:Wmass-shift} show the time
evolution of the $\rho^0$ signal, which is very noisy. Although
the signal decays quickly, a short time plateau is apparent in the
data for both clover and Wilson fermions. Similar behavior is seen
in Figs.~\ref{neutron:mass-shift} and \ref{neutron:Wmass-shift} in
the case of the neutron, for which the statistical signal is
better. (Fig.~\ref{neutron:amaze} shows the clover neutron mass
shifts out to $t=16$, and will be discussed below.) A plateau also
appears in Figs.~\ref{delta:mass-shift} and
\ref{delta:Wmass-shift} for the $\Delta^0$. The time plateaus
appear at larger time steps for the Wilson data, as would be
expected from the smaller value of the lattice spacing. The
optimal time fit ranges are a compromise between statistical
errors and systematic time evolution. Using the effective mass
shift results and chi-square values as a guide, and assuming
single exponential time fits of the ratio, $R(t)$, we fit
propagator ratio data for a given particle, mass, and electric
field across three time steps. Although we find that it is often
possible to fit at smaller times for larger $\kappa$, we prefer to
choose time plateau ranges independent of the quark mass being
studied in order to avoid introducing spurious systematic effects.

For clover data, a time plateau in the mass shift data for the
neutron, Fig.~\ref{neutron:mass-shift}, begins about time step
$5$. We take this as typical of the octet baryons and fit the
others in this same time range. The $\Delta^0$ in
Fig.~\ref{delta:mass-shift} is noisier than the neutron but also
has a plateau in the same region, evident for the heavier masses;
we fit the same time range for the decouplets as for the octets.
The $\rho^0$ in Fig.~\ref{rho:mass-shift} is even noisier than the
$\Delta^0$, but has a plateau with acceptable chi-square values
slightly further from the time origin. Our final choices for the
optimal time plateaus are $t=5-7$ for the octets and decouplet
particles, and $t=6-8$ for the $\rho^0$ and $K^{*0}$. Similar
considerations guide our choices of time plateaus in the Wilson
case, and we choose to fit time steps $14-16$ in the case of octet
baryons (see Fig.~\ref{neutron:Wmass-shift}) and $9-11$ for the
decouplets (Fig.~\ref{delta:Wmass-shift}), which again are
noisier. We also fit the $\rho^0$ (Fig.~\ref{rho:Wmass-shift}) 
and $K^{*0}$ with time steps $13-15$ and $10-12$, 
respectively. The $\Lambda^0_{s}$, the
flavor-singlet octet member, is a special case. It is much noisier
than the other octet members and we were not able to extract
results from the Wilson simulations with the same time plateau as
the other octet baryons. In the clover case, the polarizability is
the largest of all the 10 particles listed, but also with very
large error bars.

Figs.~\ref{neutron:amaze}a,b,c, and d are a striking demonstration
of the $E^2$ dependence of all of our mass shifts. These figures
show the mass shifts, Eq.(\ref{ratio}), for the neutron for all
quark masses and all four nonzero electric field values for clover
fermions. Since our nonzero electric field values differ by
factors of 2, we have rescaled the mass shifts for each of these
by a factor of 4. Clearly, the time evolution of the mass shifts
is essentially identical in these graphs - even the error bars
scale with the factor of 4. This means that quadratic dependence
on the electric field is assured for any choice of time fit range.
The other particles, including the pion, display the same type of
behavior, and this is seen for both clover and Wilson fermions.
When we fit both $E^2$ and $E^4$ coefficients, we find no evidence of
$E^4$ or higher terms and the chi-squared values on our electric
field parabola fits are quite small.

Although Fig.~\ref{neutron:amaze} leaves little doubt about the
$E^2$ dependence in our electric field range, we decreased the
field in the clover calculation by a factor of 10 for the first 20
configurations and reexamined the mass shift plots to see if the
parabola shapes changed. The shifts were 100 times smaller than
before, but the time behavior of the plots was almost identical to
the original field strength configurations for the same 20
configurations.
\begin{figure}
\begin{center}
%\epsfxsize=2.5 in
%\epsfbox{epol_vc_uudd.ps}
\includegraphics[width=5.00in]{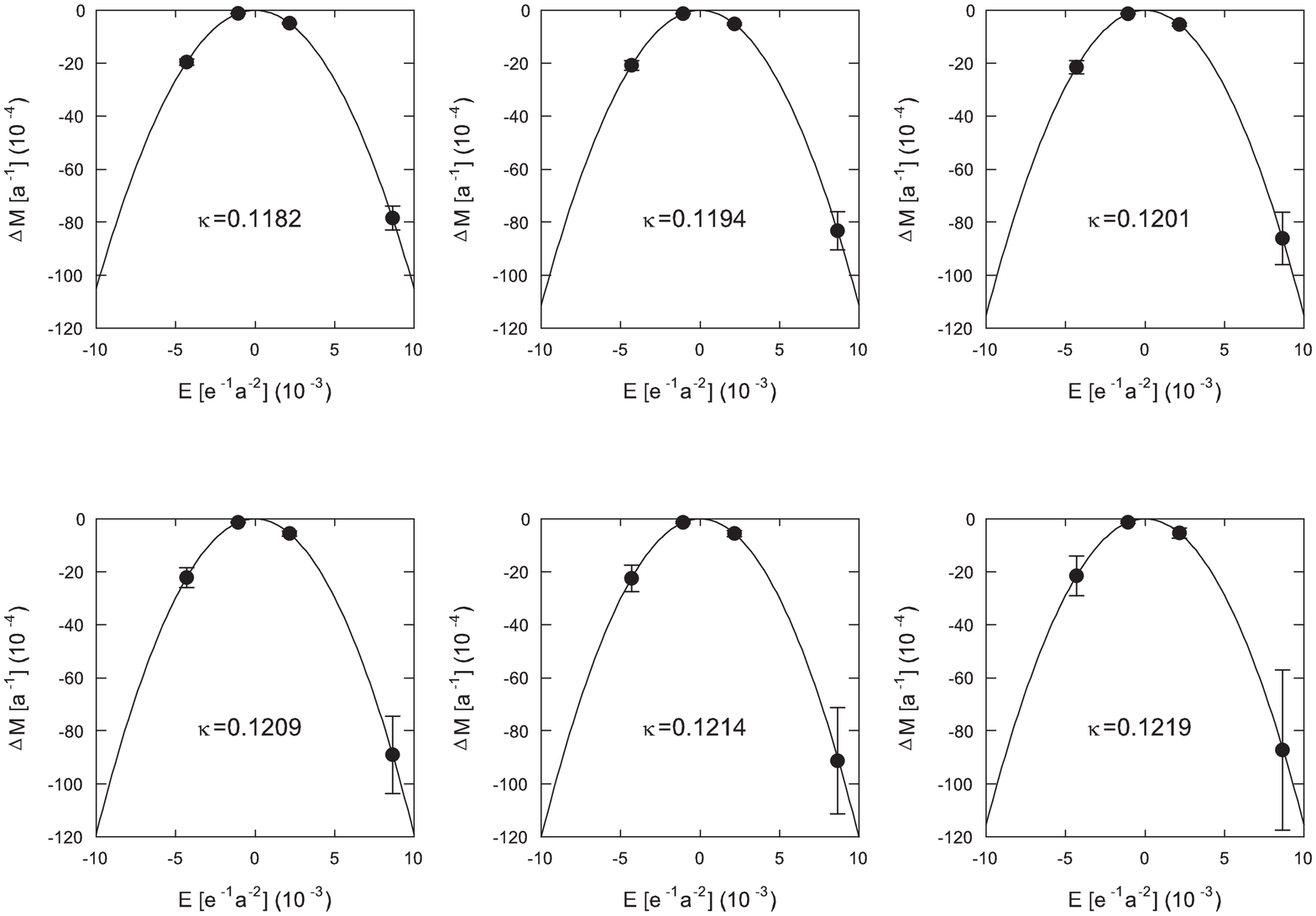}
\caption{The effective lattice mass shift, $\Delta M$ (in units of
$a^{-1}_{\rm clover}$) as a function of electric field in units of
$10^{-3}e^{-1}a^{-2}_{\rm clover}$ for the neutron for the six
values of quark mass for clover fermions. The time fit range of
propagators is 5-7.} \label{neutron-six}\vspace{.75cm}
\includegraphics[width=3.250in, angle=90]{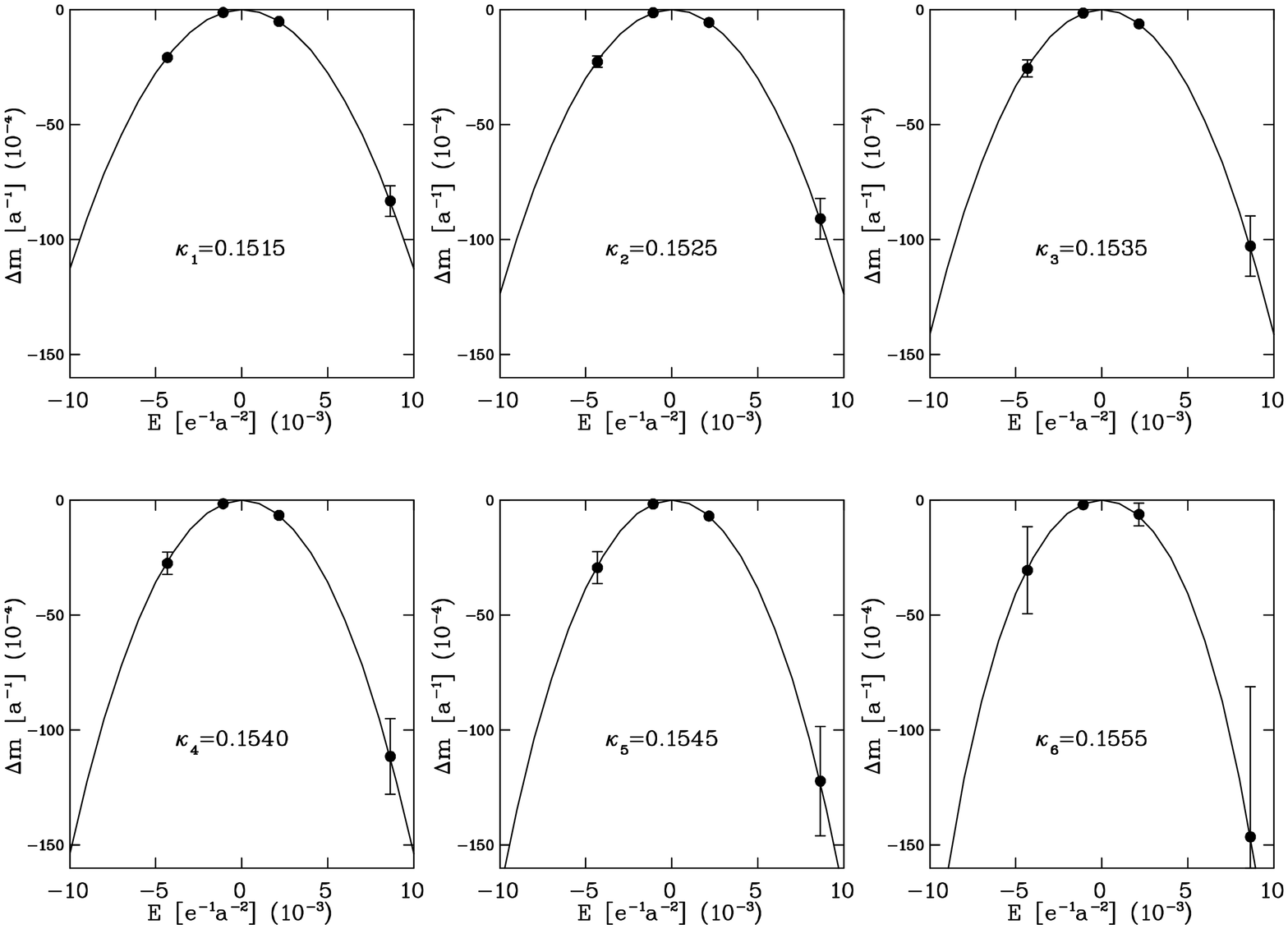}
\caption{The same as Fig.~10, but for Wilson fermions. The time
fit range of propagators is 14-16.} \label{neutron-Wsix}
\end{center}
\end{figure}

\begin{figure}
\begin{center}
\includegraphics[width=3.75in, angle=90]{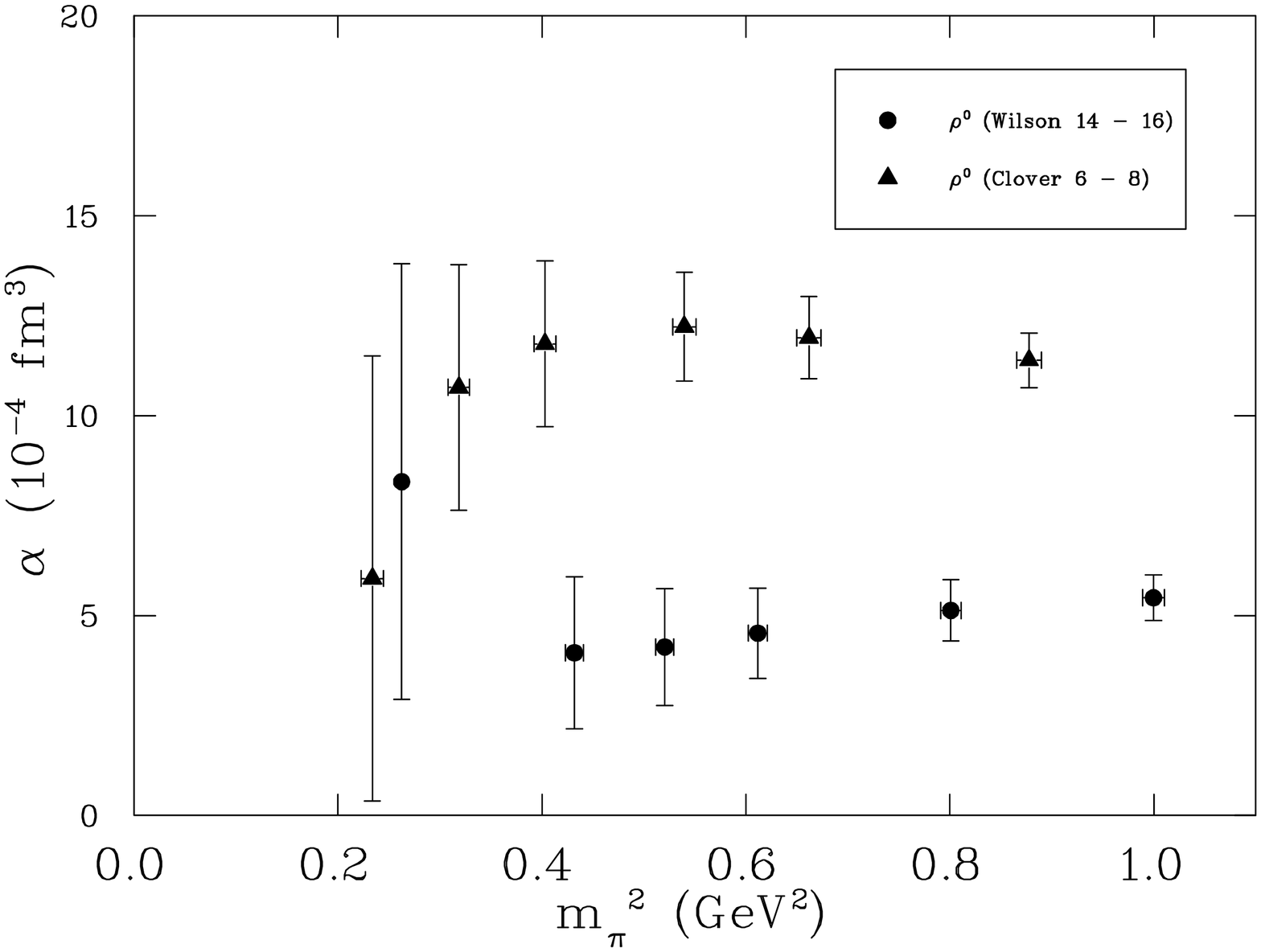}
\caption{The $\rho^0$ meson electric polarizability coefficient,
$\alpha$, in units $10^{-4}$ fm$^3$ as a function of the lattice
pion mass squared in GeV$^2$.} \label{rho:alpha}\vspace{1cm}
\includegraphics[width=3.75in, angle=90]{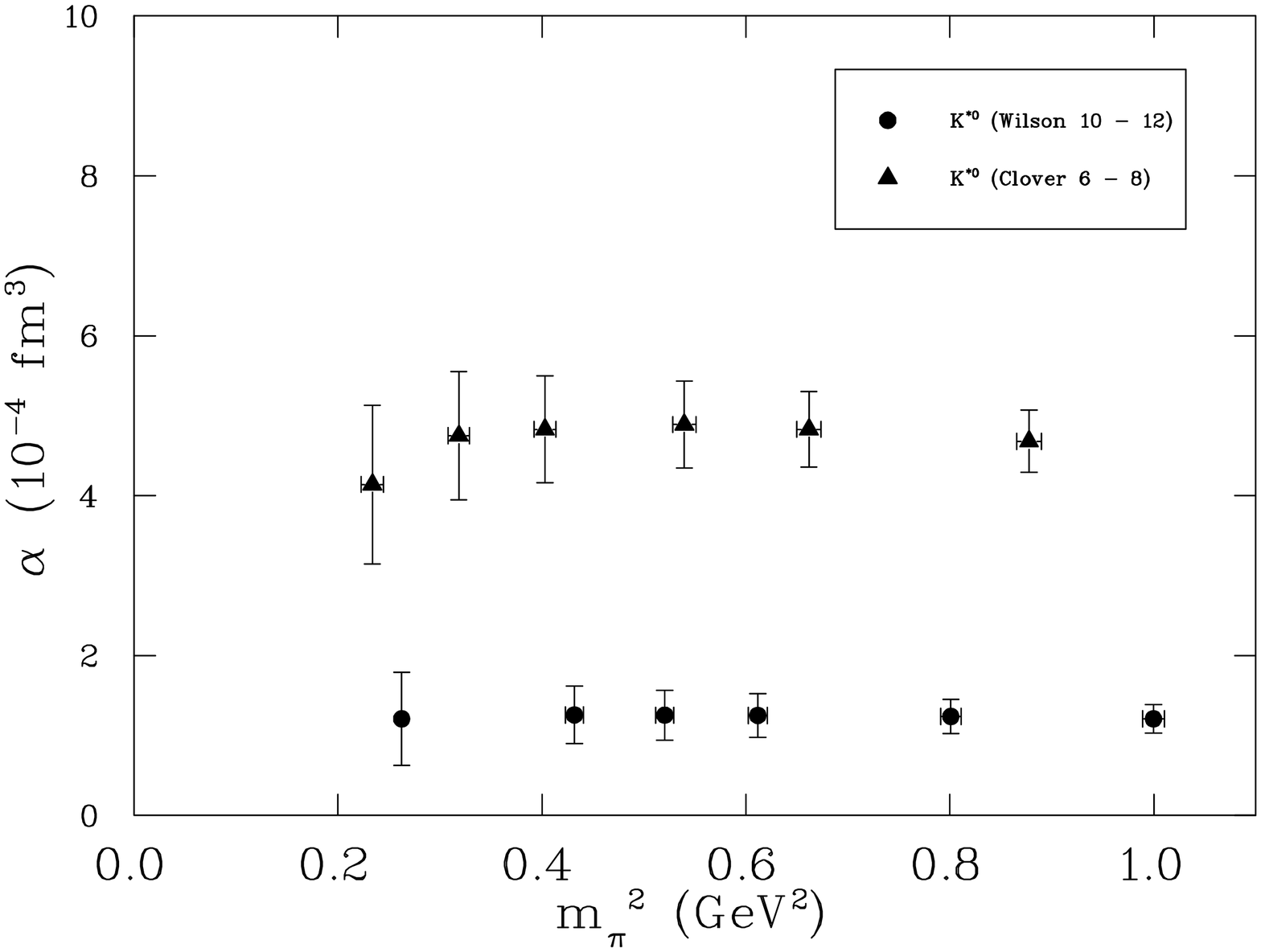}
\caption{The same as for Fig.~12, except for the $K^{*0}$.}
\label{k:alpha}
\end{center}
\end{figure}

\begin{figure}
\begin{center}
\includegraphics[width=3.75in, angle=90]{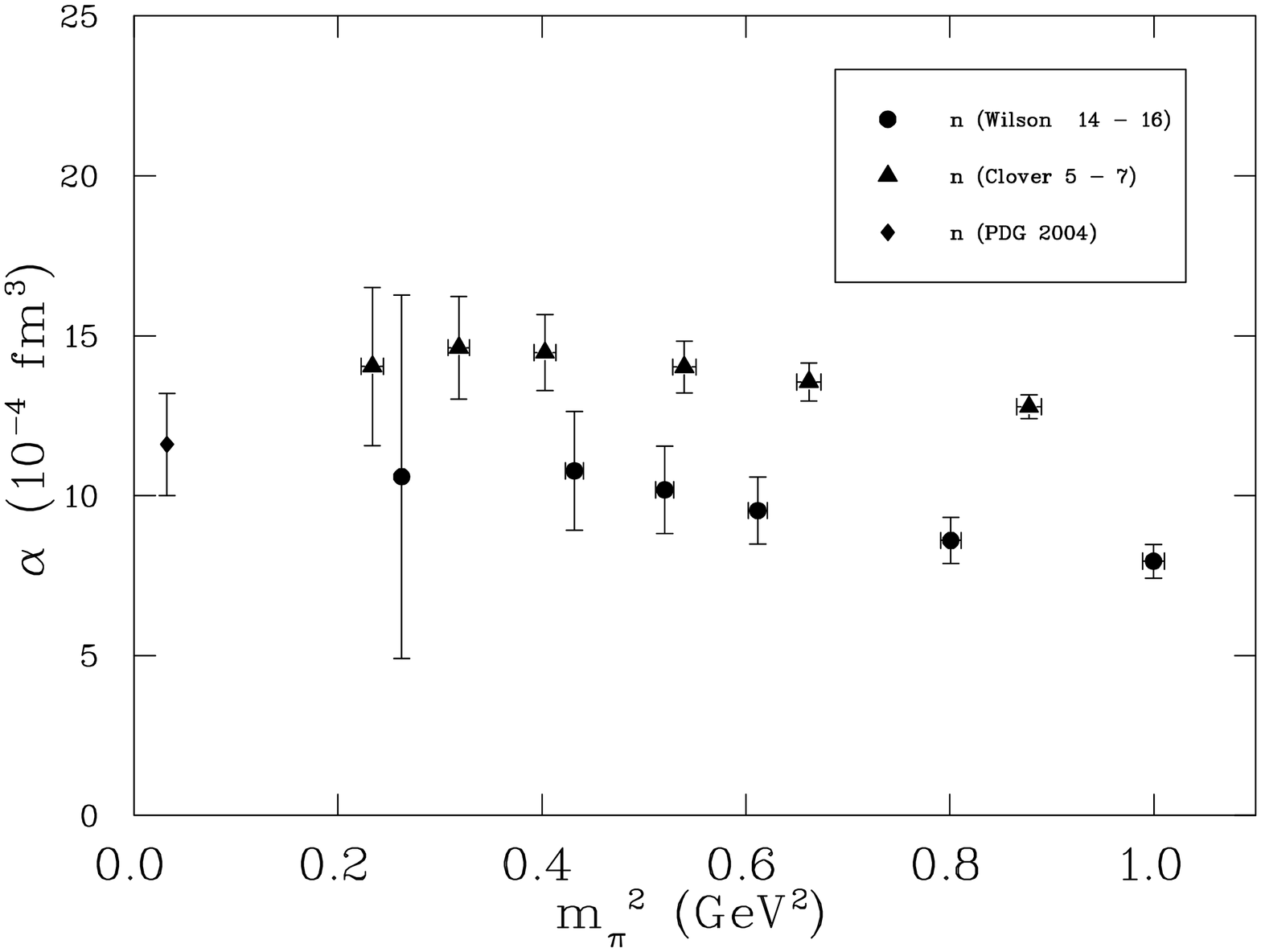}
\caption{The same as for Fig.~12, except for the neutron. The
world average is shown at the physical pion mass squared.}
\label{neutron:alpha}\vspace{1cm}
\includegraphics[width=3.75in, angle=90]{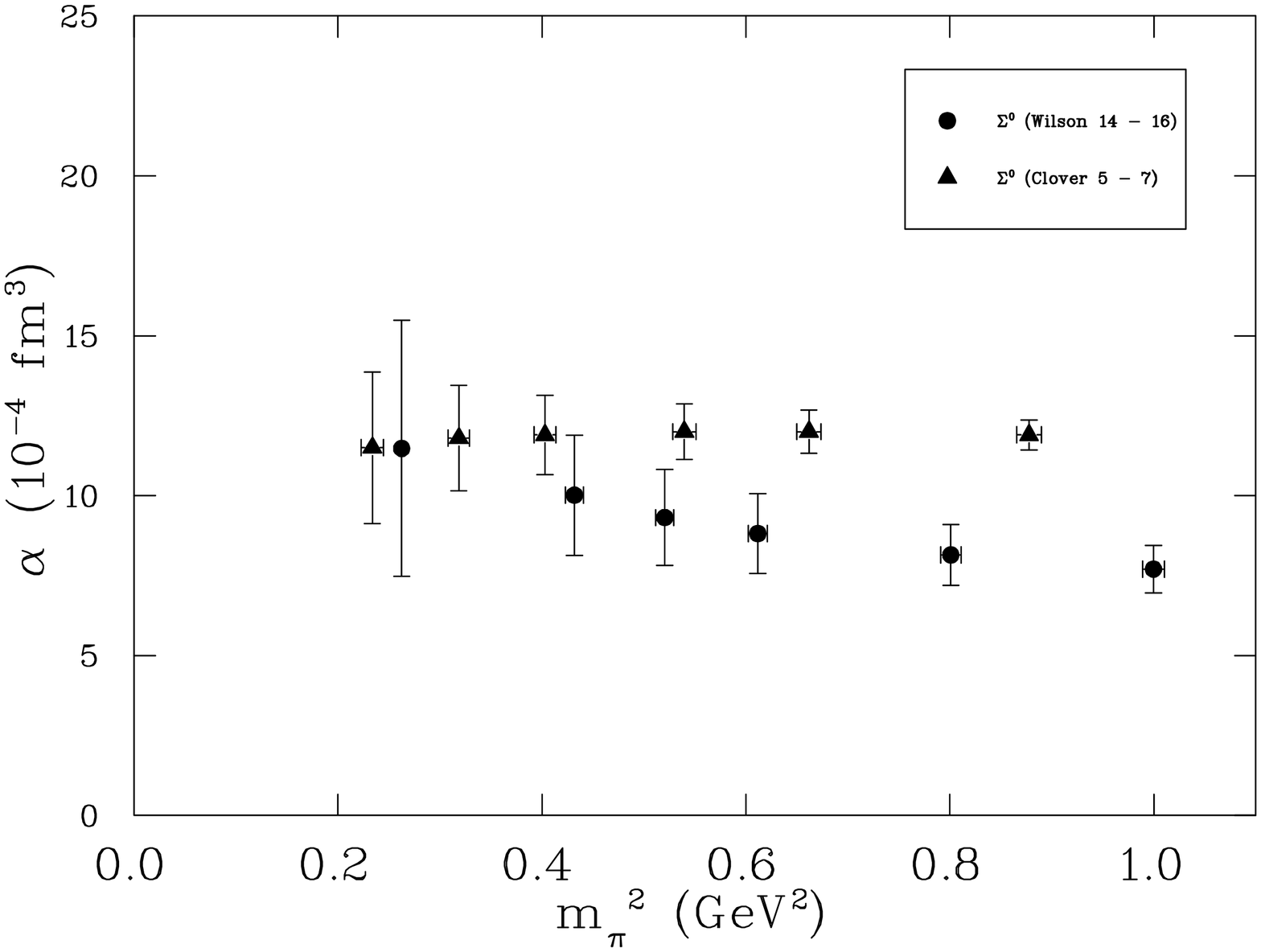}
\caption{The same as for Fig.~12, except for the $\Sigma^0$.}
\label{sigma:alpha}
\end{center}
\end{figure}

\begin{figure}
\begin{center}
\includegraphics[width=3.75in, angle=90]{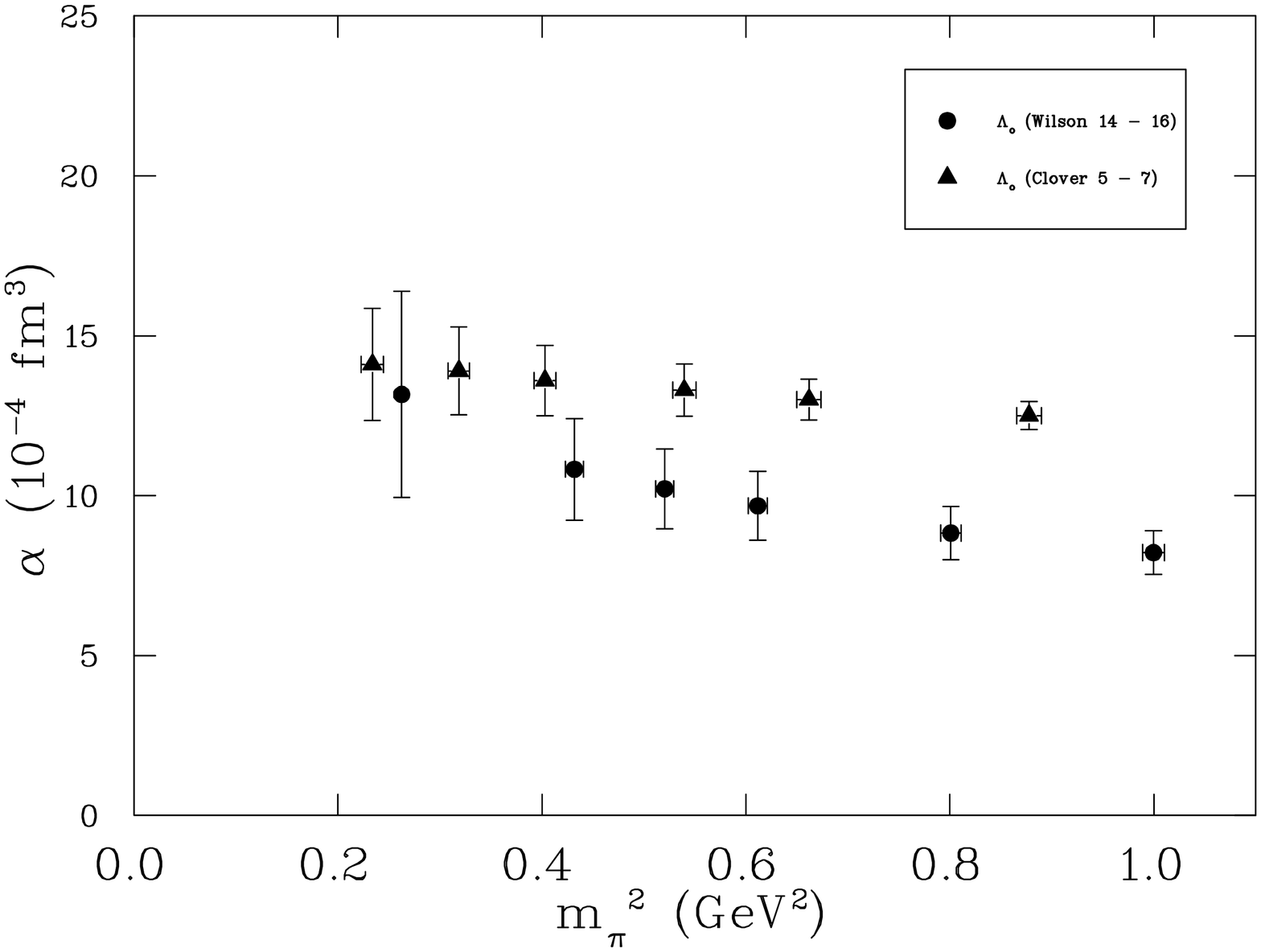}
\caption{The same as for Fig.~12, except for the $\Lambda^0_o$.}
\label{lam:alpha}\vspace{1cm}
\includegraphics[width=3.75in, angle=90]{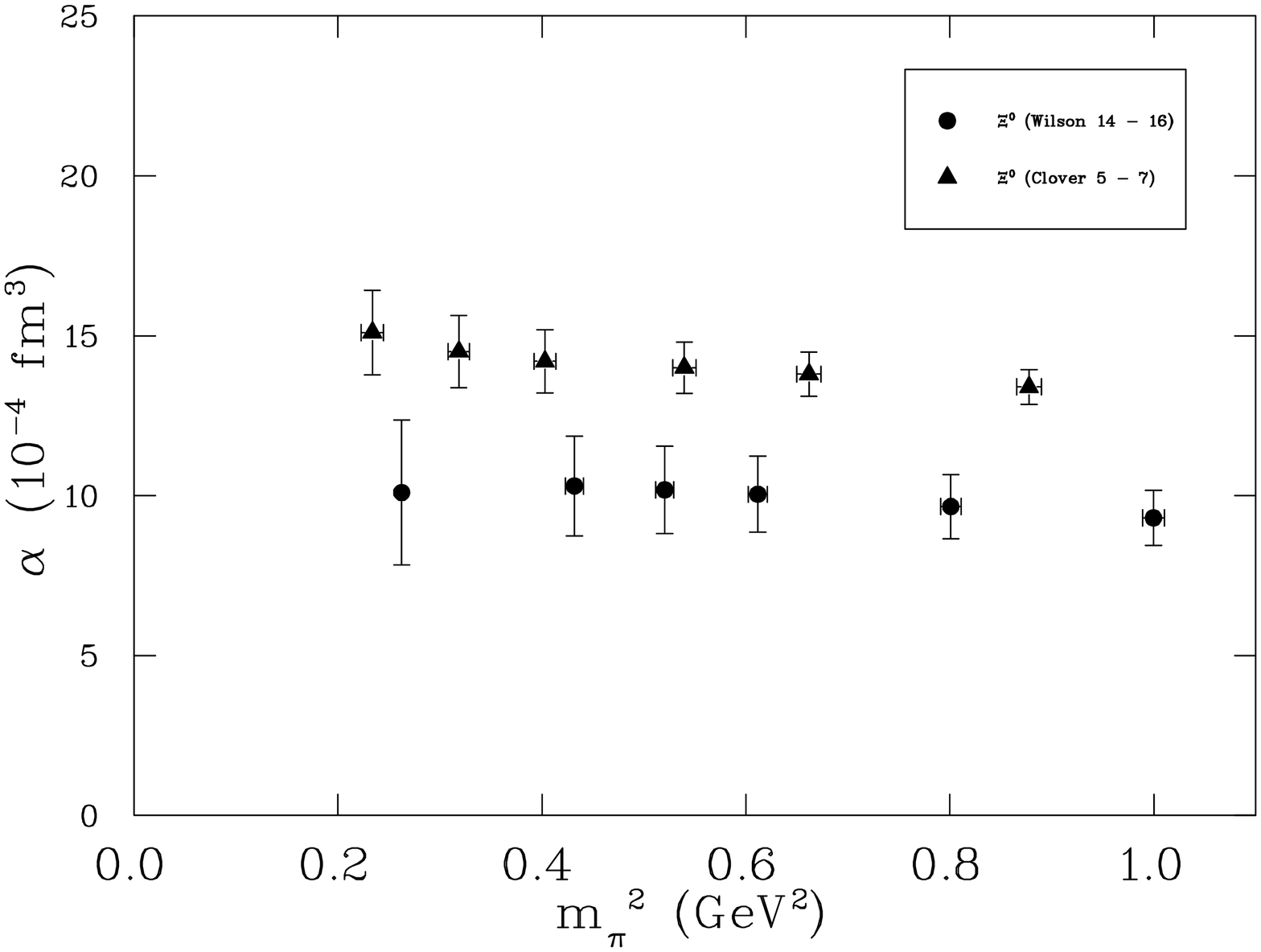}
\caption{The same as for Fig.~12, except for the $\Xi^0$.}
\label{xi:alpha}
\end{center}
\end{figure}

\begin{figure}
\begin{center}
\includegraphics[width=3.75in, angle = 90]{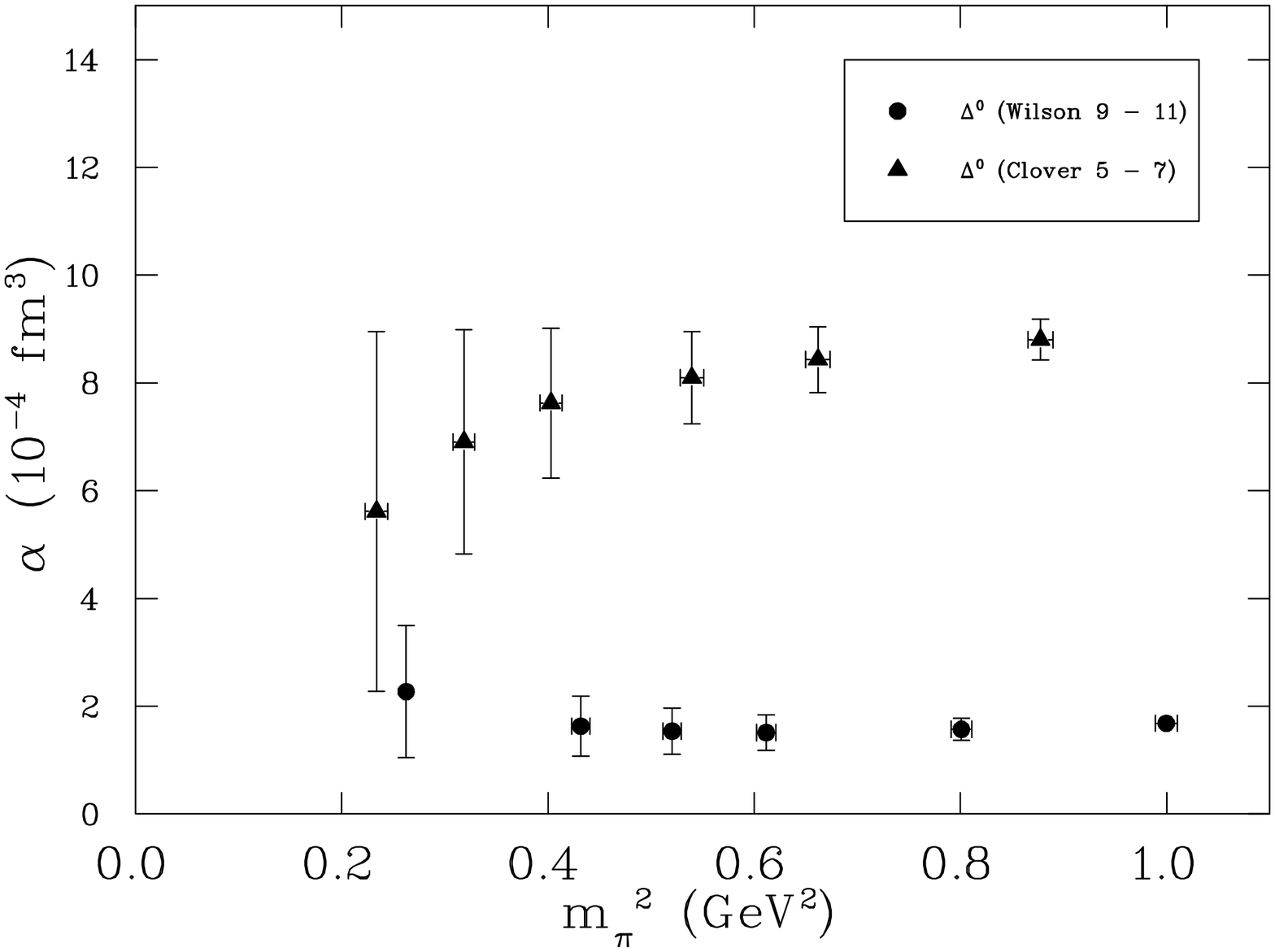}
\caption{The same as for Fig.~12, except for the $\Delta^0$.}
\label{delta:alpha} \vspace{1cm}
\includegraphics[width=3.75in, angle = 90]{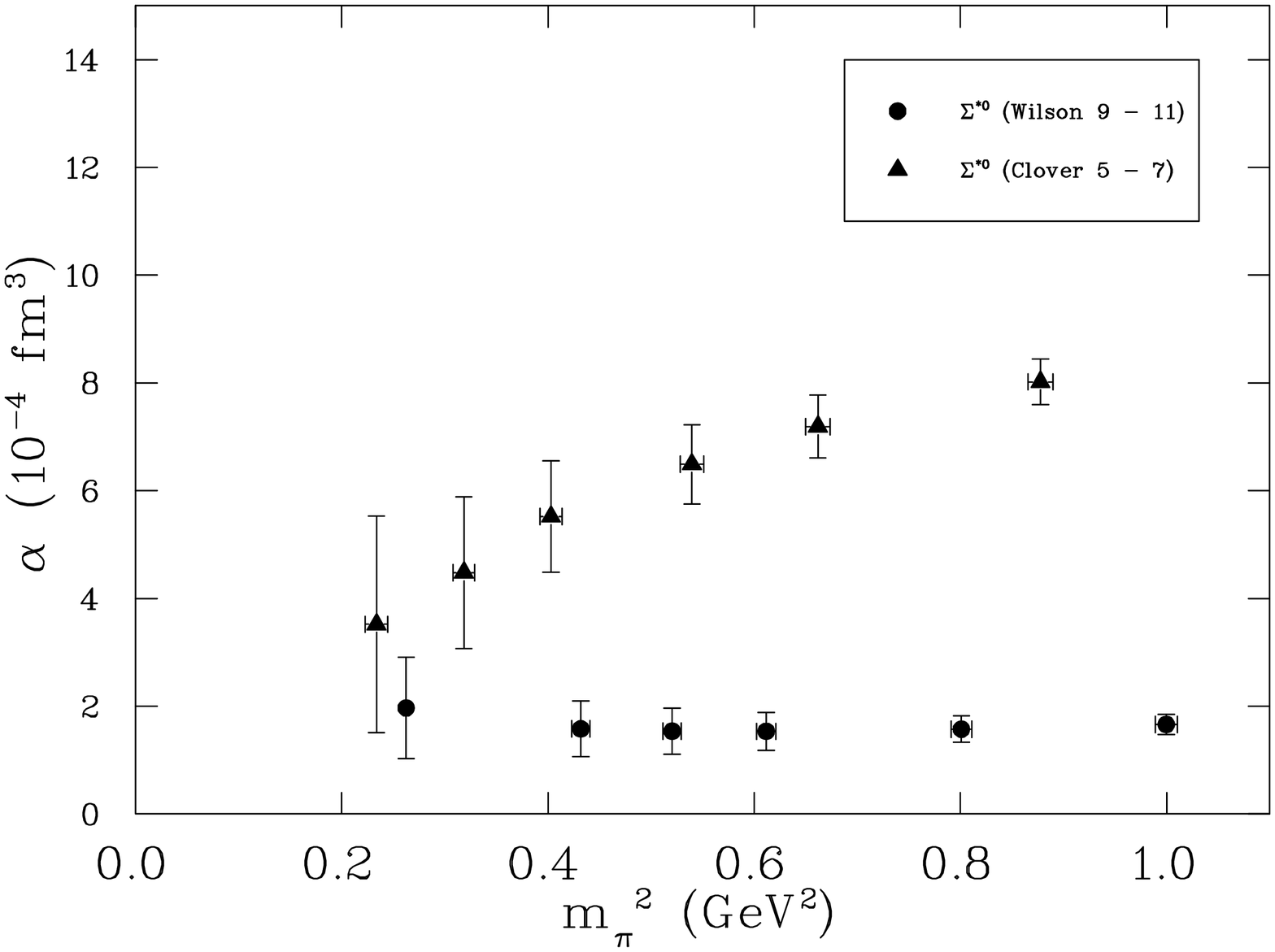}
\caption{The same as for Fig.~12, except for the $\Sigma^{*0}$.}
\label{sigma*:alpha}
\end{center}
\end{figure}

\begin{figure}
\begin{center}
\includegraphics[width=3.75in, angle = 90]{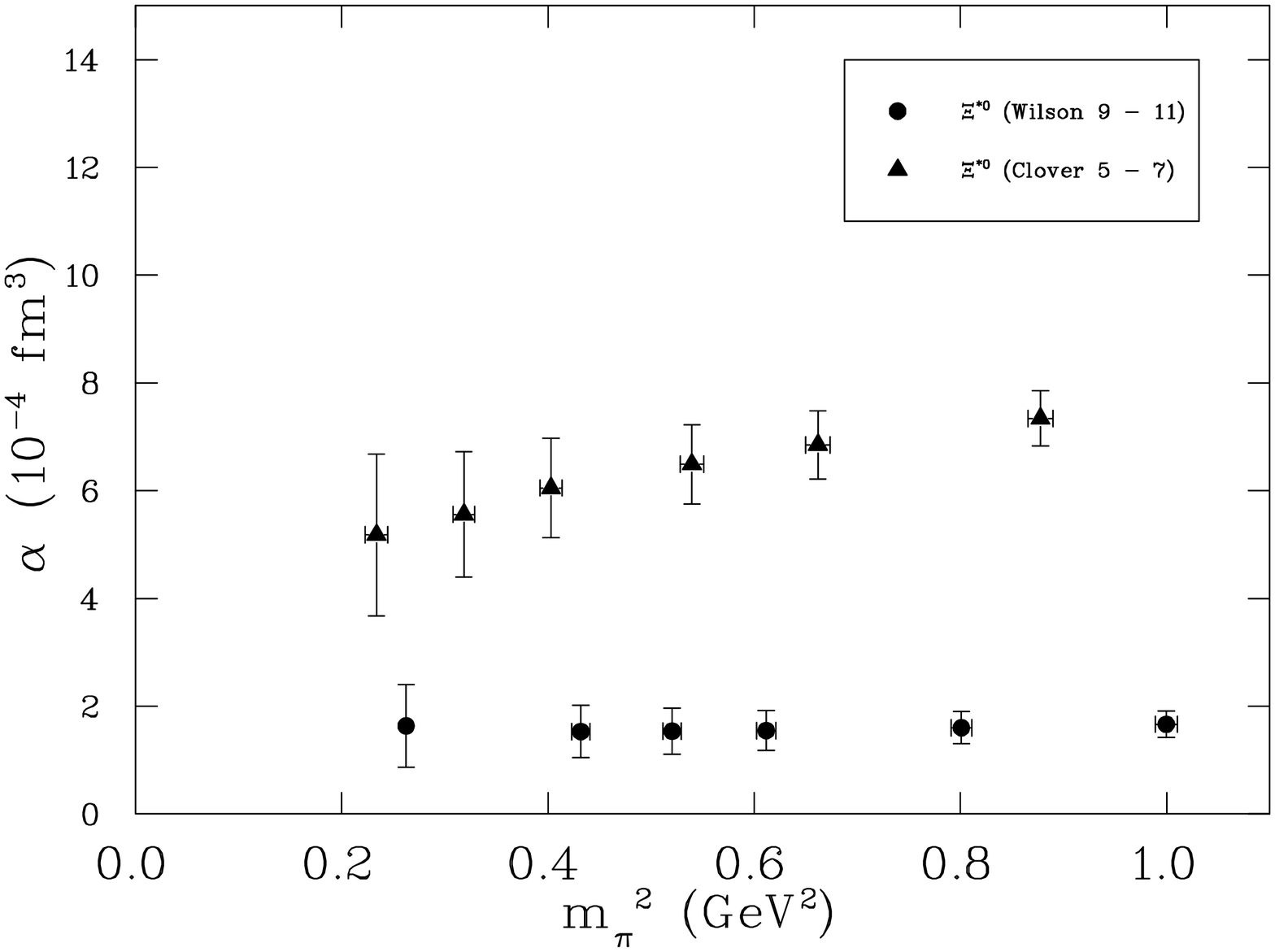}
\caption{The same as for Fig.~12, except for the $\Xi^{*0}$.}
\label{xi*:alpha}
\end{center}
\end{figure}

Figs.~\ref{neutron-six} and
\ref{neutron-Wsix} show examples of fit parabolas and error bars 
for the neutron for all six
values of quark mass. The time fit range of the propagators 
is $5-7$ in the clover case and
$14-16$ in the Wilson case. Again, Fig.~\ref{neutron:amaze}
guarantees a quadratic field dependence. The averaging procedure
described earlier over $\vec{E}$ and $-\vec{E}$ has removed odd
terms in the electric field, and we just saw that there is no
evidence of $E^4$ or higher terms in the results.
Figs.~\ref{rho:alpha}-\ref{xi*:alpha} show the pion mass
dependence of the extracted values of the electric polarizability,
$\alpha$, for both clover and Wilson fermions. We do not attempt
to extract the $\pi^0$ or $K^0$ polarizabilities because of the
problems encountered in finding a fitting plateau. Our clover
results for the other non-strange particles are very similar to
the preliminary results given in \cite{chr1}. We will discuss the
particles by groups: first the mesons, next the octet baryons, and
finally the decouplet baryons. All are plotted on the same
vertical scale ($10^{-4}$ fm$^3$) so that error bars may be
compared. Table~\ref{clovertab} tabulates the final results for
the electric polarizability coefficient for various hadrons for
clover fermions; Table~\ref{Wilsontab} gives the same results for
the Wilson case.

The results for the $\rho^0$ meson are given in
Fig.~\ref{rho:alpha}. There are definite incompatibilities in the
Wilson and clover signals at the larger pion masses. However, as
discussed in the last section, we do not necessarily expect the
Wilson and clover formulations to agree at the larger pion masses.
At the smallest pion masses, the results seem to be becoming more
compatible, although the error bars in both cases are getting
quite large. The result at our smallest quark mass is in the range
$\sim 5-10\times 10^{-4}$ fm$^3$. The results for the $K^{*0}$ are
given in Fig.~\ref{k:alpha}. In both calculations there is a
significant reduction in the polarizability coefficient for
$K^{*0}$ as compared to the $\rho^0$; the reduction is slightly
larger in the Wilson case. The results for the $K^{*0}$ do not
seem to be converging as well as for the $\rho^0$ at the smallest
quark mass. Note that the fit in the Wilson $K^{*0}$ case has been
moved 4 time steps closer to the propagator origin to achieve a
signal; this could be contributing to the larger reduction of the
Wilson results . The result here is in the range $\sim 1-4\times
10^{-4}$ fm$^3$.


The octet baryons (n, $\Sigma^{0}$, $\Lambda^0_o$, $\Xi^{0}$) are
represented by Figs.~\ref{neutron:alpha}-\ref{xi:alpha}. The
signals here are the best of all the particles for both Wilson
and clover fermions. The behavior of the all four octet members
are quite similar to one another when the Wilson or clover results
are compared among themselves. The clover results tend to be a bit
larger than the Wilson ones in all cases. All tend to move toward
larger values as the pion mass is decreased. There is a
convergence of the results for smaller masses. (The $\Xi^{0}$ in
Fig.~\ref{xi:alpha} is a possible exception.) As we will discuss
shortly, the trend of the neutron results are compatible with
experiment. All particles in
Figs.~\ref{neutron:alpha}-\ref{xi:alpha} end up with values in the
range of $\sim 10-15\times 10^{-4}$ fm$^3$ at our smallest pion
mass. We do not present a graphical result for the $\Lambda^0_s$,
but the values obtained in the clover case are given in
Table~\ref{clovertab}.

The decouplet baryons ($\Delta^0$, $\Sigma^{*0}$, $\Xi^{*0}$) are
represented by Figs.~\ref{delta:alpha}-\ref{xi*:alpha}. Here the
relative signals are not as strong as for the octets (most evident
at smaller pion mass) and the values themselves are smaller. Like
the octet case, the Wilson or clover decouplet results are quite
similar to one another when compared among themselves. While we
still have the usual incompatibilities at large pion masses, there
is a trend for the results to approach one another for the smaller
masses. (The largest disagreement is the $\Xi^{*0}$ in
Fig.~\ref{xi*:alpha}). The results at the smallest pion masses are
all on the order of $\sim 2-5\times 10^{-4}$ fm$^3$. This is quite
reduced from the octets, whose values range $\sim 10-15\times
10^{-4}$ fm$^3$ for small mass. As in the case of the $K^{*0}$
meson, it was necessary to move the time steps toward the
propagator origin in order to achieve a signal in the Wilson case;
this could contribute to the relatively large disagreements here
between Wilson and clover as opposed to the octet case.

The only experimental result available to us for comparison is the
neutron. Because of the absence of free neutron targets, actual
Compton scattering experiments must use neutrons bound in the
deuteron or other nuclei. The most recent experiments have used
either quasi-free deuteron Compton scattering($\gamma\, d
\longrightarrow \gamma\, p\, n $) or elastic deuteron scattering
($\gamma\, d \longrightarrow \gamma\, d $). The Particle Data
Group result from 2004 for the neutron is $\alpha_n =
11.6^{+1.9}_{-2.3}$~\cite{pdg}. The result of a calculation using
heavy baryon chiral perturbation theory for the neutron has been
given as $13.4\pm 1.5$~\cite{BKSM}, where the error bars are
associated with numerical integration and higher order
contributions. Our Wilson and clover results are consistent with
these values within Monte Carlo errors at our lowest pion masses,
which is very encouraging. However, more work needs to be done,
including chiral extrapolations of the present data, before more
precise comparisons can be done.

Since the polarizability coefficient scales like length cubed, it
is extremely sensitive to the assigned lattice scale, $a$.
Clearly, we must get this right if we are to extract
experimentally meaningful values for the polarizabilty
coefficient. The fact that both our clover and Wilson results for
the neutron are tending toward the experimental result at the
smallest pion masses is a strong indication of the correctness of
our methods. Turning this around, this is a excellent place to set
the lattice scale from experiment, once the neutron electric
polarizability is determined with greater accuracy.

\section{Conclusions}

In this study we have used the methods of lattice QCD to evaluate
the electric polarizability coefficient for neutral hadrons with
both Wilson and clover fermions using improved L\"usher-Weiss
gauge fields. We were able to extract and compare electric
polarizability coefficients for the vector mesons $\rho^0$ and
$K^{*0}$; the octet baryons n, $\Sigma^0$, $\Lambda_{o}^{0}$, and
$\Xi^0$; and the decouplet baryons $\Delta^0$, $\Sigma^{*0}$, and
$\Xi^{*0}$. The $\Lambda_{s}^{0}$ was extracted only in the clover
case. As a rule, the Wilson polarizability results are smaller
than the clover; however, the two calculations tend to converge at
the smallest pion masses studied. We find that the
polarizabilities of the octet baryons and of the decouplet baryons
behave as a group; all have similar mass dependencies and values
when the clover or Wilson results are compared among themselves.
This is in stark contrast to the similarity in the charge radii of
the proton and charged delta seen in Ref.~\cite{terry2}.
Apparently, the electric polarizability is not correlated to the
overall electromagnetic size of the hadron. In addition,
Ref.~\cite{terry2} sees a nontrivial squared charge radius
behavior for the neutral baryons, the neutron having a negative
charge radius and the others being zero or positive, which also
does not seem to be reflected in the polarizability. A simple
harmonic oscillator model~\cite{jackson} for charges q and -q in
an electric field gives $\alpha = 2q^2/(m\omega^2)$ (more
generally, one obtains a sum over the squared charges of the
constituents), where $m\omega^2=k$ is the spring constant.
Assuming the constitutent mass, $m$, of the quarks is a constant,
a larger energy frequency $\omega$ would explain the smaller
polarizability of the decouplets in this model.

The polarizabilities of the octet baryons show the best agreement
between clover and Wilson. Outside of the $\Lambda^0_s$, the
polarizabilities mostly tend upward as the pion mass is decreased.
They all have values in the neighborhood of $\sim 10-15 \times
10^{-4}$ fm$^{3}$ at the smallest pion mass. 
Agreement of our results with experiment
for the neutron is an encouraging sign. There is more
disagreement in the values of the decouplet baryons. Both fermion
formulations agree that the values are decreased relative to the
octets, but like the $K^{*0}$ relative to the $\rho^0$, the
reduction is larger for the Wilson case than the clover case.
Nevertheless, the clover results are tending downward toward the
Wilson values, and both end up with values in the range $2-5\times
10^{-4}$ fm$^{3}$ at the lowest pion mass. The greatest
incompatibility in the calculations appears in the meson sector,
especially the $K^{*0}$ vector meson. A possible source of the
disagreement between Wilson and clover vector mesons and decouplet
baryons seems to be the shorter time interval that was necessary
to achieve a signal for these particles in the Wilson case. We
have also seen that the pseudoscalar mesons do not have an
identifiable mass shift plateau region for either lattice fermion
formulation.

Much theoretical work remains to be done in the field of hadron
polarizability. Besides extending the present calculations to the
chiral regime, magnetic polarizabilities can be measured using
external field techniques for both charged and neutral hadrons.
Generalized polarizabilities, measured in polarized photon,
polarized proton scattering, are beginning to be studied and
measured~\cite{general}, and are candidates for lattice
calculations. There is also a need to extend chiral perturbation
theory calculations of Compton scattering to isolate the
polarizability coefficients for mesons and the other octet and
decouplet baryons so that chiral extrapolations of lattice data
may be done. In addition, the disconnected diagram contribution to
the meson polarizabilities and should be investigated, especially
in the pseudoscalar channel.

\section{Acknowledgments}

We are grateful to many people and institutions for the resources
and time necessary to carry out these investigations. WW would
like to thank the Sabbatical Committee of the College of Arts and
Sciences of Baylor University as well as the National Science
Foundation grant no.~0070836 for support. This work was also
partially supported by the National Computational Science Alliance
under grants PHY990003N, PHY890044N, PHY040021N and utilized the
Origin-2000 and the IBM p690 computers. This author also thanks
The George Washington University Physics Department for support
through their Center for Nuclear Studies Visiting Scholars
Program. In addition, this work has been supported by DOE under
grant no.~DE-FG02-95ER40907, with computer resources at NERSC and
JLAB.

\newpage
%
%%%%%%%%%%%%%%%%%%%%%%%%%%%%%%%%%%%%%%%%%%%%%%%%%%%
\begin{center}
\begin{table*}  % The * here means wide table across columns.
\caption{The electric polarizabilities from the calculation with
clover action using six $\kappa$ values. The units of the electric
polarizability are $10^{-4}$ fm$^{3}$. The pion masses were fit on
time steps $11-13$ from the propagator origin and are given in GeV.} \vspace{1.0cm}
\label{clovertab}
\begin{tabular}{lccccccc}
\hline \hline $\kappa$ & 0.1182 & 0.1194 & 0.1201 & 0.1209 &
0.1214 & 0.1219 &        \\ $ m_\pi$  & $0.937\pm .006$\quad  &
\quad $0.814\pm .007$\quad  & \quad $0.735\pm .008$\quad  & \quad $0.635\pm .008$\quad  &
\quad $0.564\pm .009$\quad  & \quad $0.483\pm .011$ &
\\ \hline \hline Mesons & & & & & & & fit range \\ $\rho^0$ &
 11.4 $\pm$ .7 &
 12.0 $\pm$ 1.0 &
 12.2 $\pm$ 1.4 &
 11.8 $\pm$ 2.0 &
 10.7 $\pm$ 3.0&
 5.9 $\pm$ 5.6 &
 6 to 8  \\
  $K^{*0}$ &
 4.7 $\pm$ .4  &
 4.8 $\pm$ .5  &
 4.9 $\pm$ .5  &
 4.8 $\pm$ .7  &
 4.8 $\pm$ .8  &
 4.1 $\pm$ 1.0  &
 6 to 8 \\
\hline \hline \multicolumn{8}{l}{Baryon octet} \\ n   &
  12.8 $\pm$ .4 &
  13.6 $\pm$ .6 &
  14.0 $\pm$ .8 &
  14.5 $\pm$ 1.2 &
  14.6 $\pm$ 1.6 &
  14.0 $\pm$ 2.5 &
   5 to 7  \\$\Sigma^0$ &
  11.9 $\pm$ .5 &
  12.0 $\pm$ .7 &
  12.0 $\pm$ .9 &
  11.9 $\pm$ 1.2 &
  11.8 $\pm$ 1.6 &
  11.5 $\pm$ 2.4 &
   5 to 7  \\ $\Lambda^0_o$ &
  12.5 $\pm$ .4 &
  13.0 $\pm$ .6 &
  13.3 $\pm$ .8 &
  13.6 $\pm$ 1.1 &
  13.9 $\pm$ 1.4 &
  14.1 $\pm$ 1.7 &
   5 to 7 \\ $\Lambda^0_s$ &
  30 $\pm$ 1 &
  32 $\pm$ 21 &
  40 $\pm$ 24 &
  52 $\pm$ 35 &
  55 $\pm$ 43 &
  53 $\pm$ 86 &
   5 to 7  \\ $\Xi^0$      &
  13.4 $\pm$ .5 &
  13.8 $\pm$ .7 &
  14.0 $\pm$ .8 &
  14.2 $\pm$ 1.0 &
  14.5 $\pm$ 1.1 &
  15.1 $\pm$ 1.3 &
   5 to 7 \\
\hline \hline \multicolumn{8}{l}{Baryon decuplet} \\ $\Delta^{0}$&
  8.8 $\pm$ .4 &
  8.4 $\pm$ .6 &
  8.1 $\pm$ .9  &
  7.6 $\pm$ 1.4 &
  6.9 $\pm$ 2.1 &
  5.6 $\pm$ 3.3 &
  5 to 7 \\ $\Sigma^{*0}$&
  8.0 $\pm$ .4 &
  7.2 $\pm$ .6 &
  6.5 $\pm$ .7  &
  5.5 $\pm$ 1.0 &
  4.5 $\pm$ 1.4 &
  3.5 $\pm$ 2.0 &
  5 to 7  \\ $\Xi^{*0}$   &
  7.3 $\pm$ .5 &
  6.8 $\pm$ .6 &
  6.5 $\pm$ .7  &
  6.1 $\pm$ .9 &
  5.6 $\pm$ 1.2 &
  5.2 $\pm$ 1.5 &
  5 to 7  \\ \hline \hline
\end{tabular}
\end{table*}
\end{center}

%
%%%%%%%%%%%%%%%%%%%%%%%%%%%%%%%%%%%%%%%%%%%%%%%%%%%
\begin{center}
\begin{table}  % The * here means wide table across columns.
\caption{The electric polarizabilities from the calculation with
Wilson action using six $\kappa$ values. The units of the electric
polarizability are $10^{-4}$ fm$^{3}$. The pion masses were fit on
time steps $11-13$ from the propagator origin and are given in GeV.} \vspace{1.0cm}
\label{Wilsontab}
\begin{tabular}{llcccccc}
\hline \hline $\kappa$ & 0.1515 & 0.1525 & 0.1535 & 0.1540 &
0.1545 & 0.1555 &    \\ $ m_\pi$ & $1.000\pm .005$\quad  &
\quad $0.895\pm .006$\quad  & \quad $0.782\pm .006$\quad  & \quad $0.721\pm .006$\quad  & \quad
$0.657\pm .007$\quad   & \quad $0.512\pm .007$  &
\\ \hline \hline Meson & & & & & & & fit range \\ $\rho^{0}$&
   5.0  $\pm$     0.5 &
   4.8  $\pm$     0.7 &
   4.4  $\pm$     0.9 &
   4.0  $\pm$     1.2 &
   3.5  $\pm$     1.6 &
   2.8  $\pm$     4.8 &
   13-15        \\
   $K^{*0}$ &
   1.2 $\pm$   0.2 &
   1.2 $\pm$   0.2 &
   1.3 $\pm$   0.3 &
   1.3 $\pm$   0.3 &
   1.3 $\pm$   0.4 &
   1.2 $\pm$   0.6 &
   10-12  \\
\hline \hline \multicolumn{8}{l}{Baryon octet}  \\ n            &
   7.9  $\pm$     0.5 &
   8.6  $\pm$     0.7 &
   9.5  $\pm$     1.0 &
   10.2 $\pm$     1.4 &\
   10.8 $\pm$     1.9 &
   10.6 $\pm$     5.7 &
   14-16     \\
$\Sigma^0$   &
   7.7  $\pm$     0.7 &
   8.1  $\pm$     0.9 &
   8.8  $\pm$     1.3 &
   9.3  $\pm$     1.5 &
   10.0 $\pm$     1.9 &
   11.5 $\pm$     4.0 &
   14-16     \\
$\Lambda^0_o$    &
   8.2  $\pm$     0.7 &
   8.8  $\pm$     0.8 &
   9.7  $\pm$     1.1 &
   10.2 $\pm$     1.2 &
   10.8 $\pm$     1.6 &
   13.2 $\pm$     3.2 &
   14-16     \\
$\Xi^0$      &
   9.3  $\pm$     0.9 &
   9.7  $\pm$     1.0 &
   10.0 $\pm$     1.2 &
   10.2 $\pm$     1.4 &
   10.3 $\pm$     1.6 &
   10.1 $\pm$     2.3 &
   14-16     \\
\hline \hline \multicolumn{8}{l}{Baryon decuplet}
\\ $\Delta^{0}$&
   1.7  $\pm$     0.1 &
   1.6  $\pm$     0.2 &
   1.5  $\pm$     0.3 &
   1.5  $\pm$     0.4 &
   1.6  $\pm$     0.6 &
   2.3  $\pm$     1.2 &
   9-11       \\
$\Sigma^{*0}$  &
   1.7  $\pm$     0.2 &
   1.6  $\pm$     0.2 &
   1.5  $\pm$     0.4 &
   1.5  $\pm$     0.4 &
   1.6  $\pm$     0.5 &
   2.0  $\pm$     0.9 &
   9-11           \\
$\Xi^{*0}$     &
   1.7  $\pm$     0.2 &
   1.6  $\pm$     0.3 &
   1.5  $\pm$     0.4 &
   1.5  $\pm$     0.4 &
   1.5  $\pm$     0.5 &
   1.6  $\pm$     0.8 &
   9-11       \\
\hline \hline
\end{tabular}
\end{table}
\end{center}
%%%%%%%%%%%%%%%%%%%%%%%%%%%%%%%%%%%%%%%%%%%%%%%%%%%
%

\end{document}